\documentclass[11pt,twoside]{article}

\usepackage{asp2004}
\usepackage{epsf}
\usepackage{epsfig}
\usepackage{lscape}
\usepackage{amsmath,amssymb}

\begin{document}
\title{The Hanle Effect in Atomic and Molecular Lines: \\A New Look at
the Sun's Hidden Magnetism}


\author{J. Trujillo Bueno,$^{1,2}$ A. Asensio Ramos,$^{1}$ and 
N. Shchukina$^{3}$}
\affil{%
$^1$Instituto de Astrof\'{\i}sica de Canarias, V\'{\i}a L\'actea s/n, 
E-38205 La~Laguna, Tenerife, Spain\\ \vspace{2pt}
$^2$Consejo Superior de Investigaciones Cient\'{\i}ficas, Spain\\
\vspace{2pt}
$^3$Main Astronomical Observatory, National Academy of Sciences, 
Zabolotnogo 27, 03680 Kyiv, Ukraine}

\begin{abstract}
This paper\footnote{This article combines in one
contribution Trujillo Bueno's invited keynote paper  
and the contributed papers by Asensio Ramos \& Trujillo Bueno 
and by Shchukina \& Trujillo Bueno.}
reviews some of the most recent advances in the application of the 
Hanle effect to solar physics, and how these developments 
are allowing us to explore 
the magnetism of the photospheric regions that look ``empty'' in 
solar magnetograms---that is, the Sun's ``hidden'' magnetism. 
In particular, we show how a joint analysis of the Hanle effect in 
atomic and molecular lines indicates that there is a vast amount 
of hidden magnetic energy and unsigned magnetic flux localized 
in the (intergranular) downflowing regions of the quiet solar 
photosphere, carried mainly by tangled fields at sub-resolution 
scales with strengths between the equipartition field values and 
$\sim1$\,kG.
\end{abstract}


\vspace{-8pt}
\section{Introduction}

At any given time during the solar magnetic activity cycle
most of the solar surface is covered by the internetwork regions 
of the ``quiet'' Sun. Although such regions look ``empty'' 
(i.e., devoided of magnetic signatures) in low-resolution 
magnetograms, high-spatial-resolution observations reveal the 
presence of mixed magnetic polarities with a mean unsigned flux 
density of the order of 10\,G, when choosing the best compromise 
between the polarimetric sensitivity and the spatio-temporal resolution 
currently attainable 
\citep*[e.g.,][]{t2 LR99,t2 DC03,t2 Kh03,t2 LS04,t2 MG06}.
However, when these Zeeman-effect polarization signals are 
interpreted considering that the magnetic field is not spatially 
resolved (for example, by assuming that one or more magnetic 
components coexist with a non-magnetic component within the 
spatio-temporal resolution element of the observations), it is 
found that the filling factor of the magnetic component(s) is $\sim1$\%. 

The disadvantage of the Zeeman effect as a diagnostic tool is that the 
amplitudes of the measured polarization signals become smaller 
with the increasing degree of cancellation of mixed magnetic 
polarities within the spatio-temporal resolution element. Therefore, 
vanishing Zeeman polarization does not necessarily imply absence
of magnetic fields.
On the contrary, the photospheric plasma of the quiet Sun is expected to 
be permeated by highly tangled field lines with resulting mixed 
magnetic polarities on very small spatial scales, well beyond the 
current spatial resolution limit 
\citep[e.g.,][]{t2 St94,t2 Ca99,t2 SA03a,t2 SN03}.
For this reason, it is believed that Zeeman-effect diagnostics 
with the present-day instrumentation is showing us only
the ``tip of the iceberg'' of solar surface magnetism ($\sim1$\% 
of the photospheric volume).  

How can we reliably investigate the magnetism of the remaining 
$\sim99$\% of the quiet solar atmosphere? 
How can we obtain empirical information on the distribution of 
mixed-polarity magnetic fields at sub-resolution scales 
that are ``hidden'' to the Zeeman effect?
Do such fields carry most of the unsigned magnetic flux 
and magnetic energy of the Sun, or is this flux and/or the magnetic 
energy dominated by small-scale magnetic flux concentrations in 
the kG range, as some recent investigations based on the Zeeman
effect of the \ion{Fe}{i} lines at 6301.5\,\AA\ and 6302.5\,\AA\ 
seem to suggest? How is the magnetic energy of the quiet Sun 
distributed between upflows and downflows? Is the total magnetic 
energy stored in the quiet internetwork regions greater or 
smaller than that of the kG fields of the super-granulation 
network? Is this energy significant enough to compensate for 
the radiative losses of the outer solar atmosphere? 
How is energy transported and dissipated in a mixed 
magnetic-polarity environment? Does this hidden magnetic 
flux complicate the topology of the magnetic fields of the outer 
solar atmosphere, or there simply are magnetic canopies, with 
mainly horizontal fields, in the quiet regions of the solar 
chromosphere? 

Since the Zeeman effect as a diagnostic tool is ``blind'' to 
magnetic fields that are tangled on scales too small to be 
resolved,\footnote{This applies to the spectral line 
polarization induced by the Zeeman effect, but not to the Zeeman 
broadening of the intensity profile.} the investigation of the 
magnetism of the quiet Sun via spectro-polarimetry must rely on other
tools as well. In addition to Zeeman diagnostics developed 
to interpret the observed {\em asymmetries} of Stokes 
profiles---resulting from the co-existence of opposite magnetic 
polarities that do not cancel completely within the resolution element
\citep*[e.g.,][]{t2 SA96,t2 SS02},
and/or from hyperfine-structure effects 
\citep*[e.g.,][]{t2 LA02},
we also need to apply ``new'' diagnostic techniques based on physical 
mechanisms whose observable signatures do not suffer from 
cancellation effects. 

Two such mechanisms are the Hanle effect and the line broadening by the
Zeeman effect \citep[e.g.,][]{t2 St94}.
We argue that the Hanle effect can be suitably {\em complemented} 
with the Zeeman broadening of the intensity profiles of some 
carefully selected near-IR lines, as a tool
to obtain information on the distribution of tangled magnetic
fields at sub-resolution scales in the quiet regions of the solar photosphere. In 
particular, we will review in some detail how the most recent 
advances in the application of the Hanle effect to 
solar physics has led us to the conclusion that the hidden 
magnetic fields of the quiet Sun 
carry a vast amount of unsigned flux and energy.  

In Sect.~2 we highlight some of the most recent investigations 
on the magnetism of the quiet Sun based on the interpretation of 
Zeeman polarization signals.
In Sect.~3 we consider the Hanle 
effect in the \ion{Sr}{i} 4607\,\AA\ line, showing how the 
observed scattering polarization amplitudes suggest the presence of 
a tangled magnetic field at sub-resolution scales, with a mean field 
strength $\langle B \rangle \sim 100$\,G. In Sect.~4 we address the 
question of whether a significantly smaller $\langle B \rangle$ 
could perhaps be inferred by assuming a magnetic field topology 
different from that of an isotropic microturbulent field. 
In Sect.~5 we discuss the {\em direct} observational evidence of 
magnetic depolarization provided by other spectral lines, such as 
the multiplet No.\,42 of \ion{Ti}{i}. 
In Sect.~ 6 we review the work done on the Hanle effect in molecular 
lines, pointing out some key questions (e.g., the fact that the 
inferred $\langle B \rangle \sim10$\,G), whose answer led us to the 
conclusion that the strength of the hidden field ``fluctuates''
on the spatial scales of solar granulation, with rather weak fields 
above the granular regions, but with a distribution of stronger 
fields in the intergranular regions (capable of producing saturation 
of the Hanle effect in the \ion{Sr}{i} 4607\,\AA\ line formed there).
Finally, in Sect.~7 we summarize the main conclusions and discuss 
possible lines of future research.

\section{Diagnostics of Zeeman-Effect Polarization}

Over the last few years the observational and theoretical evidence of
small-scale mixtures of weak and strong fields in the quiet Sun has
increased considerably 
\citep*[e.g.,][]{t2 LR99,t2 SL00,t2 Li02,t2 SS03,t2 Kh03,t2 SL04,t2
SA03a,t2 SA03b,t2 Ca99,t2 SN03,t2 Vo03},
but so has the controversy about the true abundance of kG fields in the
internetwork regions of the quiet Sun 
(e.g., \citealt{t2 DC03,t2 BC03,t2 LS04,t2 SN04,t2 SA03b,t2 SA04,
t2 Wi05,t2 Kh05,t2 MG06}; Collados, this conference).
For example, \citet{t2 DC03}
interpret the systematic difference between the flux densities 
measured in the \ion{Fe}{i} 6301.5/6302.5\,\AA\ lines
as evidence for small-scale magnetic structures with intrinsic 
field strengths larger than 1\,kG, occupying 2\% of the solar surface. 
This would imply an abundance of kG fields in the internetwork regions 
of the Sun significantly larger than that inferred by \citet{t2 Kh03}
from measurements of the Stokes profiles in the \ion{Fe}{i} lines 
at 15648\,\AA\ and 15652\,\AA. 

Concerning proxy magnetometry, it is 
important to mention that \citet{t2 Wi05}
measured a surface density of internetwork bright points (IBP) of
0.02\,Mm$^{-2}$ in sharp G-band images obtained with the Dutch 
Open Telescope, and concluded that the observed IBPs seem to outline 
mesogranular cell-like structures. 
\citet{t2 SA04}
reported instead that in their best G-band image (which contained a 
small network patch) obtained with the 1-m Swedish Solar Telescope, 
the density of bright points is 0.3\,Mm$^{-2}$, which led 
them to conclude that detected bright points cover 0.7\% of the 
solar surface. 

Recently, \citet{t2 Kh06}
has suggested that the application of Stenflo's (\citeyear{t2 St73})
line-ratio technique to the observed Stokes-$V$ profiles produced 
by the Zeeman effect in the \ion{Fe}{i} 6301.5/6302.5\,\AA\ lines 
could overestimate the fraction of (internetwork) quiet Sun 
occupied by field strengths $B>1$\,kG (see also Figs.~1 and 2 in 
\citealt{t2 KC06}), because these lines do not have exactly the 
same absorption coefficient or the same behavior in the dynamic 
and highly inhomogenous solar atmosphere \citep{t2 ST01}.
This fact is illustrated in the left panels of Fig.~1, which 
shows results of three-dimensional (3D), non-LTE radiative transfer 
calculations. 
Interestingly, the right panels of Fig.~1 suggest that 
the application of the line-ratio technique to the 
\ion{Fe}{i} 15648/15652\,\AA\ lines could indeed be a reliable 
strategy. A similar calculation for the \ion{Fe}{i} 5250.21/5247.05\,\AA\ 
lines reveals a behavior for $H_{\rm VIS}$(red) and 
$H_{\rm VIS}$(blue) (with $\Delta\lambda=\pm30$\,m\AA)
more or less similar to that shown in the left panels of Fig.~1, 
but with $H_{\rm VIS}^{5250}-H_{\rm VIS}^{5247}\approx0$ km. 
It is also important to note that while 
$S_{\rm line}\approx B_{\nu}$ for the 5250.21/5247.05\,\AA\ lines, 
$(S_{\rm line}/B_{\nu})_{6301.5}<(S_{\rm line}/B_{\nu})_{6302.5}<1$
\citep[see Fig.~8 of][right panel]{t2 ST01}.

\begin{figure}[!t]
\centering
\includegraphics[width=6.2cm]{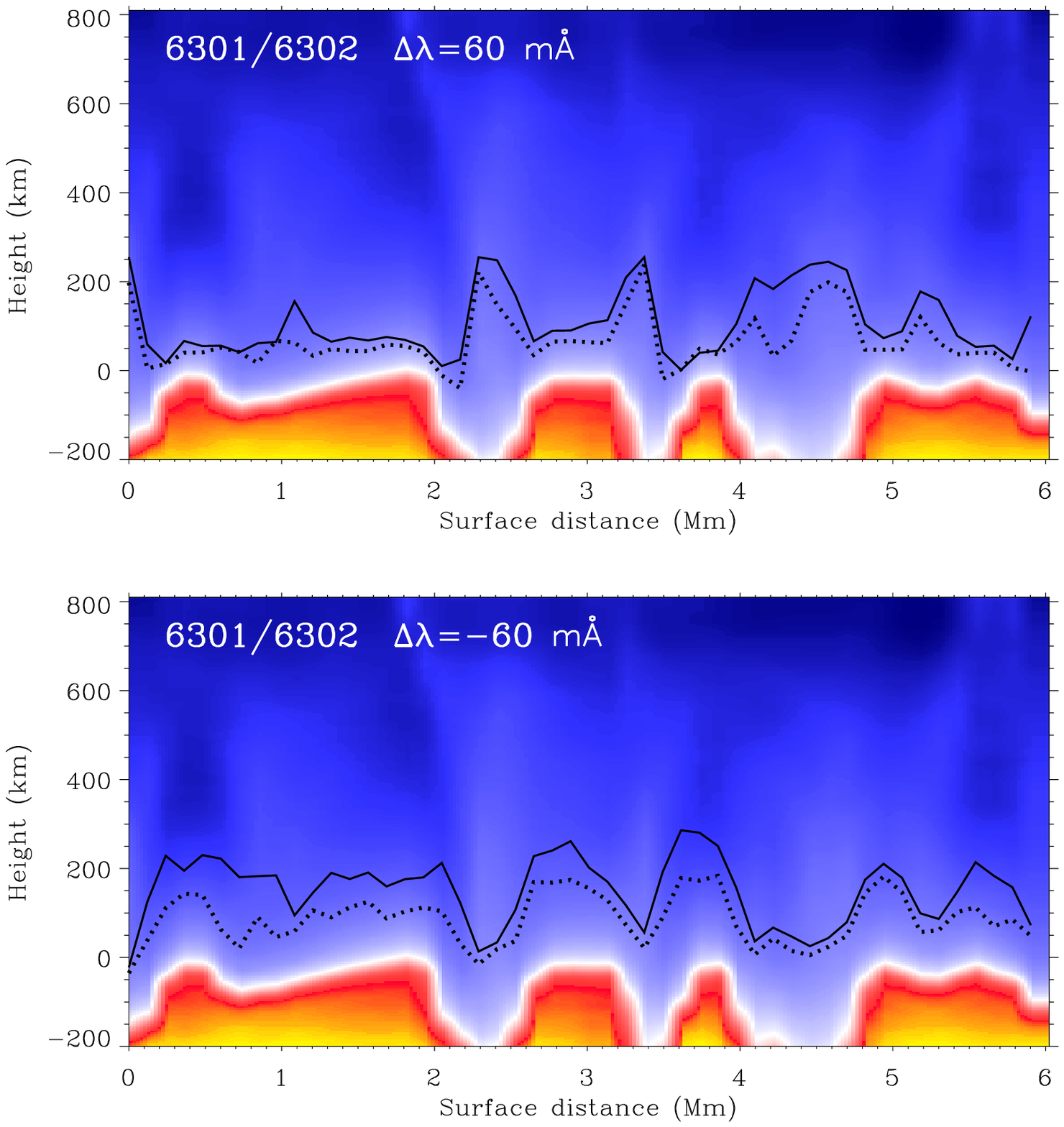}\hspace{5pt}
\includegraphics[width=6.2cm]{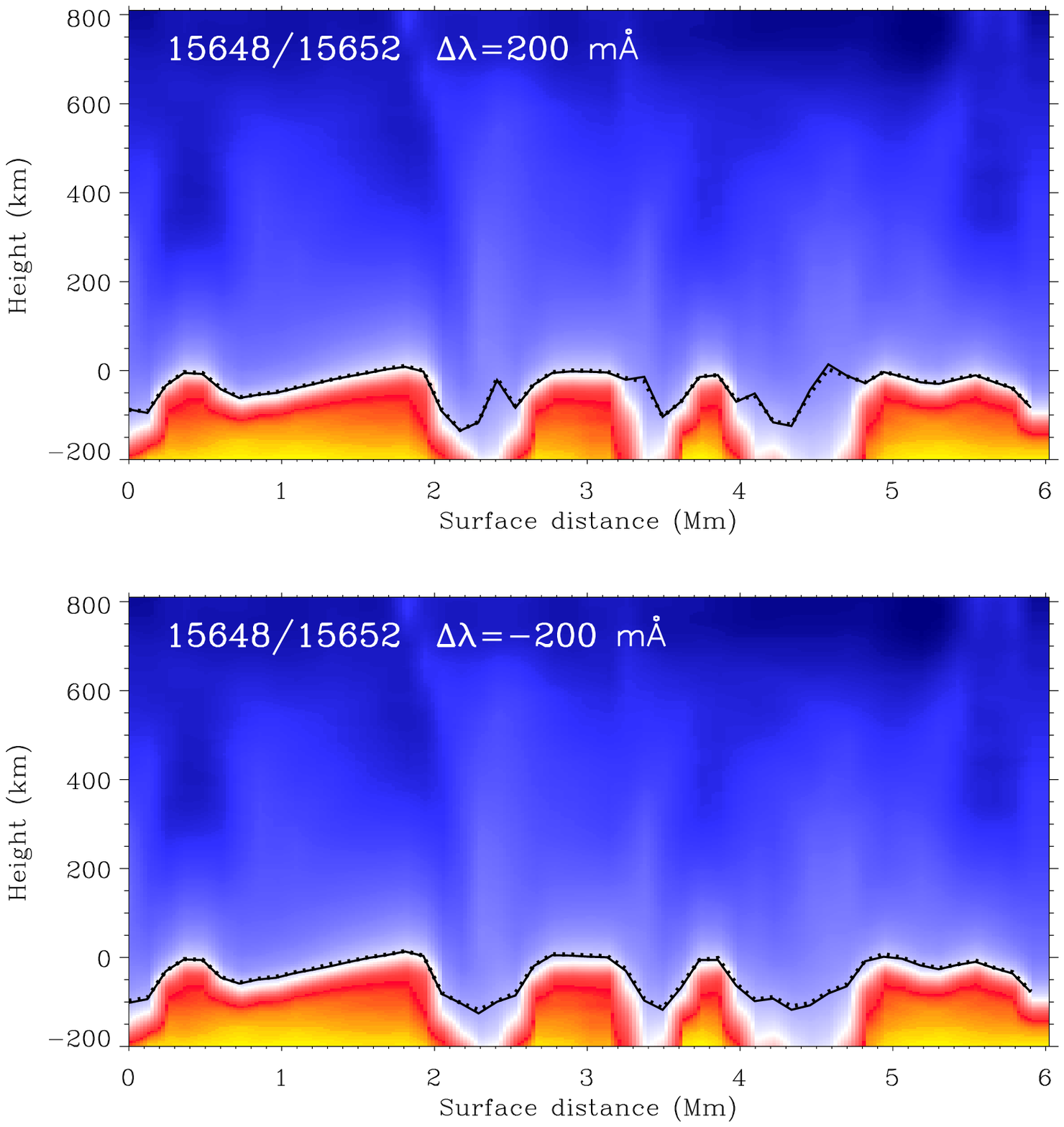}
\caption{Plots of the horizontal fluctuation of the height 
$H_{\rm line}$ at which
$\tau_{\rm line}(\Delta\lambda)=1$, for different lines and values of
$\Delta\lambda$, in a simulated observation at disk-center using a 
realistic 3D hydrodynamical model of the solar photosphere. 
The adopted values for $\Delta\lambda$ are typical distances from 
line center at which internetwork Stokes-$V$ profiles of the selected 
lines have their ``red'' and ``blue'' lobes.
{\em Left:} \ion{Fe}{i} 6301.5\,\AA\ (solid line) and 6302.5\,\AA\ 
(dotted line), for $\Delta\lambda=+60$\,m\AA\ (top panel) and 
$\Delta\lambda=-60$\,m\AA\ (bottom panel). 
Note that $H_{\rm VIS}^{6301.5}-H_{\rm VIS}^{6302.5}\approx100$\,km,
which is a significant height difference. Note also that
in the (intergranular) downflowing regions 
$H_{\rm VIS}({\rm blue})\ll H_{\rm VIS}({\rm red})$, 
while in the (granular) upflowing regions 
$H_{\rm VIS}({\rm blue})\gg H_{\rm VIS}({\rm red})$.
{\em Right:} \ion{Fe}{i} 15648\,\AA\ (solid line) and 15652\,\AA (dotted
line), for $\Delta\lambda=+200$\,m\AA\ (top panel) and 
$\Delta\lambda=-200$\,m\AA\ (bottom panel).
Note that $H_{\rm IR}^{15648}-H_{\rm IR}^{15652}\approx0$\,km,
and that $H_{\rm IR}({\rm blue})\approx H_{\rm IR}({\rm red})$ 
in the granular and intergranular regions, independently. 
Finally, it is also important to note that 
$H_{\rm VIS}>H_{\rm IR}$.}
\label{t2 fig:Fig1}
\end{figure}

Therefore, it is at present unclear whether or not the small-scale 
magnetic fields that we ``see'' in the internetwork regions of the 
Sun via the longitudinal Zeeman effect in the 6301.5/6302.5\,\AA\ 
lines (whose filling factor is $\sim 1$\%) are sufficiently strong 
to carry a truly significant fraction of the unsigned magnetic flux 
and magnetic energy of the Sun. In this respect, we note that 
\citet{t2 Ke94}
already applied Stenflo's (\citeyear{t2 St73})
magnetic line-ratio technique to measurements of the Stokes-$V$ 
profiles of the \ion{Fe}{i} 5250.21/5247.05\,\AA\ lines,
and concluded that the strength of solar internetwork fields at the 
height of line formation is below 1\,kG with a high probability. 
Obviously, {\em multiline} observations with high spatial resolution 
and polarimetric sensitivity are urgently needed.

\section{The Hanle Effect in the Sr\,{\sc i} 4607\,\AA\ Line}

An unresolved, tangled magnetic field tends to reduce the line scattering
polarization amplitudes with respect to the zero magnetic field case. 
This so-called Hanle effect has the required diagnostic
potential for investigating tangled magnetic fields at sub-resolution scales 
\citep{t2 St82}.
The critical magnetic field strength (measured in G) that is sufficient 
to produce (via the upper-level Hanle effect) a significant change in 
the line's scattering polarization amplitude, when the excitation of 
the atomic or molecular system is dominated by radiative transition, is 
\citep[e.g.,][]{t2 TB01a}
\begin{equation}
B_{\rm H}={\frac{1.137\times10^{-7}}{t_{\rm life}\,g}}\, ,
\end{equation}
where $g$ and $t_{\rm life}$ are, respectively, 
the Land\'e factor and the radiative lifetime (measured in seconds)
of the upper level of the spectral line under consideration 
(e.g., $B_{\rm H}\approx23$\,G for the \ion{Sr}{i} 4607\,\AA\ line).

The main problem with choosing the Hanle effect as a diagnostic 
tool of the quiet Sun magnetism is how to apply it to obtain reliable 
information, given that it relies on a comparison between the 
observed linear polarization and that calculated
for the zero-field reference case. \Citet{t2 ST03}
have shown that the particular one-dimensional (1D) approach 
applied by \citet{t2 FS95} and \citet{t2 Fa01}
to the scattering polarization
observed in the \ion{Sr}{i} 4607\,\AA\ line
yields artificially low values for the strength of the 
hidden field---that is, $\langle B \rangle\approx20$\,G, 
as shown by the dashed line in the left panel of Fig.~3 in 
\citet{t2 ST03}. 
This is because the scattering polarization amplitudes calculated by 
\citet{t2 FS95} and \citet{t2 Fa01}
for the zero-field reference case, $(Q/I)_{B=0}$,  
was seriously underestimated due to their choice of the free 
parameters of ``classical" stellar spectroscopy (that is, micro- and
macro-turbulence for line broadening), which led to a sizable error 
in the value of the Hanle depolarization factor 
${\cal D}=(Q/I)/(Q/I)_{B=0}$, where $Q/I$ is the observed 
polarization amplitude. 

Note that for the currently used model of a 
microturbulent field, the depolarization factor
${\cal D}$ is {\em similar} to the factor ${\cal H}^{(2)}$ given by 
Eq.~(A-16) of \citet{t2 TM99},
with ${\cal H}^{(2)}=1$ for $B=0$\,G and ${\cal H}^{(2)}=1/5$ for 
$B>B_{\rm satur}$ (where $B_{\rm satur}\sim 10\,B_{\rm H}$ 
is the saturation field of the Hanle effect for the spectral line 
under consideration). To understand why 
${\cal D}\approx{\cal H}^{(2)}$ it suffices to note that
Eq.~(4) of \citet{t2 TB03a} 
leads to the following Eddington-Barbier
approximation for the emergent $Q/I$ at the
line-core of the \ion{Sr}{i} 4607\,\AA\ line,
\begin{equation}
Q/I\approx\frac{3}{2\sqrt{2}}(1-\mu^2)
\frac{{\cal H}^{(2)}}{1+{\delta}_u}\,{\cal A}\, ,
\end{equation}
where ${\mu}=\cos{\theta}$ (with $\theta$ the angle between 
the solar radius vector through the observed point and the 
line-of-sight), 
${\delta}_u=D^{(2)}\,t_{\rm life}\approx D^{(2)}/A_{ul}$ is 
basically the upper-level 
rate of (depolarizing) elastic collisions in units of the 
Einstein $A_{ul}$ coefficient, and
${\cal A}=J^2_0/J^0_0$ is the degree of anisotropy of the 
spectral line radiation \citep[e.g.,][]{t2 TB01a}.
The particular 1D radiative transfer fitting approach of 
\citet{t2 FS95} and \citet{t2 Fa01}
gives ${\cal D}\approx0.8$ when using the accurate observational 
data shown in Fig.~2. 

\begin{figure}[!t]
\centering
\includegraphics[width=8.9cm]{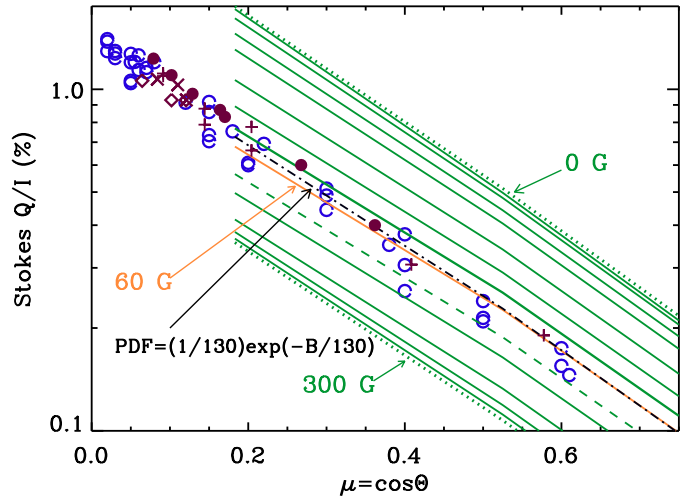}
\caption{
Center-to-limb variation 
of the fractional linear polarization
in the center of the \ion{Sr}{i} 4607\,\AA\ line after subtraction of the
continuum polarization. 
($\circ$) various observations taken by \citet{t2 St97}
during a minimum of the solar cycle.  The remaining symbols 
correspond to observations taken during the most recent maximum 
of the solar cycle: 
($\diamond$) observations obtained by \citet{t2 TB01b}; 
($\times$ and $+$) observations taken by 
\citet{t2 BM02} 
and by \citet{t2 Bo05}, 
respectively; 
($\bullet$) observations obtained in collaboration with 
M.~Bianda. Colored (or solid, dashed, and dotted) 
lines show the results of our 3D scattering polarization calculations in the 
presence  of a volume-filling and single value microturbulent
field (from top to bottom: 0, 5, 10, 15, 20, 30, 
40, 50, 60, 80, 100, 150, 200, 250 and 300\,G).
Note that there is no evidence of a significant modulation
of the observed $Q/I$ with the solar magnetic activity cycle, and 
that the best average fit to the observations is obtained for 60\,G.
The black, dashed-dotted line indicates the resulting $Q/I$ amplitudes
for the case of a microturbulent field described by an exponential 
Probability Distribution Function
with $\langle B \rangle=130$\,G. (From \citealt*{t2 TB04}.)}
\label{t2 fig:Fig2}
\end{figure} 

\Citet*{t2 TB04} 
pointed out that reliable Hanle-effect diagnostics can be achieved by means of
3D multilevel scattering polarization calculations 
in snapshots taken from realistic simulations of solar surface
convection \citep[see][]{t2 As00}.
These radiation hydrodynamical 
simulations of the photospheric physical conditions are very convincing
because spectral synthesis of a multitude of iron lines shows remarkable
agreement with the observed spectral line 
profiles \citep[e.g.,][]{t2 ST01}.
Therefore, we have used snapshots taken from such time-dependent 
hydrodynamical simulations in order
to calculate the emergent Stokes profiles via 
3D scattering polarization calculations using 
realistic multilevel atomic models.
Our first target was the \ion{Sr}{i} line at 4607\,\AA,
which is a normal triplet transition with 
$A_{ul}\approx2\times10^8\,{\rm s}^{-1}$
and Land\'e factor $g_{J_u}=1$. We point out that our synthetic 
intensity profiles (which take into account the Doppler shifts of 
the convective flow velocities in the 3D model) are  
in excellent agreement with the observations when the meteoritic 
strontium abundance is chosen. However, 
we find that the spatially and temporally averaged emergent Stokes 
profiles for $B=0$\,G give a $Q/I$ that is substantially larger 
than observed, thus indicating the need for invoking magnetic 
depolarization. As can be easily deduced from Fig.~2, we obtain 
${\cal D}\approx 0.4$.

If our 3D result for ${\cal D}$
is correct, then any analysis based on 
${\cal H}^{(2)}\approx{\cal D}\approx0.8$ (i.e., based 
on the ${\cal D}$ values obtained when the accurate $Q/I$ measurements 
of Fig.~2 are normalized to the $(Q/I)_{B=0}$ estimate of Faurobert and coworkers) would imply a mean 
strength of the hidden field substantially smaller than the actual 
value (e.g., the simplified analyses carried out by 
\citealt*{t2 SA03a} and \citealt{t2 SH03},
which were based on the estimation of ${\cal D}$ that
\citealt{t2 FS95} obtained from older observations). Obviously, 
the only way to arrive at a solid conclusion 
concerning the unsigned magnetic flux and magnetic energy density 
carried by the {\em hidden field} is to use the correct value of the
depolarization factor, ${\cal D}$.

We think that at present the 3D analysis of the observed 
scattering polarization in the \ion{Sr}{i} 4607\,\AA\ line
carried out by \citet{t2 TB04} provides the most accurate estimation of ${\cal D}$, not only because 
of the fact already mentioned that our synthetic intensity 
profiles (which take into account the Doppler shifts of the 
convective flow velocities in the 3D model) are already in 
excellent agreement with the observed intensity profiles {\em without} 
having to use the free parameters of classical stellar spectroscopy, 
but also because we carefully computed the radiation field's 
anisotropy in the 3D hydrodynamical model 
\citep[e.g., Fig.~1 of][]{t2 ST03}.
We solved the statistical equilibrium equations for the elements of 
the atomic density matrix and the Stokes-vector transfer equations
using efficient radiative transfer methods \citep{t2 TB03b}, 
and adopting realistic collisional depolarizing rate values 
\citep{t2 FS95}, 
which turn out to be the largest rates among
those found in the literature.
It is also important to point out that the recent work of 
\citet{t2 Bo05} 
strongly supports 
the result for ${\cal D}$ of \citet{t2 TB04}.

We may thus conclude that the depolarization required to fit the
$Q/I$ observations of the \ion{Sr}{i} 4607\,\AA\ line is 
with high probability of
magnetic origin.\footnote{We are currently investigating whether or not 
bound-free transitions make any significant depolarizing influence on the 
atomic alignment of the upper level of the \ion{Sr}{i} 4607\,\AA\ line 
\citep[cf.][]{t2 TB03a}. 
For the moment we point out that the largest bound-free cross sections 
correspond to the triplet levels, while the upper level of the 
\ion{Sr}{i} 4607\,\AA\ line is a singlet.} For the moment we have 
assumed that the magnetic field is isotropic and microturbulent, 
which we consider as a reasonable approximation for estimating 
the mean strength of the ``hidden'' field.\footnote{We point out 
that above a height of 300\,km in the quiet solar atmosphere
the mean-free-path of the \ion{Sr}{i} 4607\,\AA\ line photons 
increases rapidly beyond 100\,km, and that at moderate spatial resolution 
the observed Stokes $U\approx0$ 
\citep[see Fig.~1 of][]{t2 TB01b}.} 
It is obvious that with a single spectral line we do not have enough
information to constrain the shape of the Probability Distribution
Function, ${\rm PDF}(B)$, describing the fraction of 
quiet Sun occupied by magnetic fields with strength $B$.
For this reason, we chose the functional form of the PDF,
like others have also done
\citep*[e.g.,][]{t2 FS95,t2 Fa01,t2 Bo05,t2 DC06}.
The key point, however, is to
be conservative in the choice of the functional form of the PDF, 
in order to avoid exaggerating the resulting mean strength of the 
hidden field. For this reason, although we carried out calculations 
for several functional forms of the PDF, \citet{t2 TB04}
presented results only for the two forms: (a) 
${\rm PDF}(B)={\delta}(B-\langle B \rangle)$ and 
(b)~${\rm PDF}(B)={\rm e}^{-B/\langle B \rangle}/\langle B \rangle$, 
where $\langle B \rangle$ is the mean field strength.

The idealized model corresponding to
case (a) will clearly underestimate $\langle B \rangle$, but 
it is interesting to compare the ensuing magnetic energy with 
that corresponding to the kG fields of the network patches. The 
much more realistic case (b) is suggested by numerical experiments 
of turbulent dynamos and magnetoconvection 
\citep[e.g.,][]{t2 Ca99,t2 SN03,t2 Vo03}.
The ``surface'' PDF in some of these numerical experiments tends 
to be instead a {\em stretched exponential}, which, to a first 
approximation, can be fitted by a Voigt function whose tail extends 
further out into the kG range than an exponential.\footnote{In the 
turbulent dynamo experiments of \citet{t2 Ca99},
even if the most probable field strength in the interior of 
the computational box is not exactly zero, the ensuing PDF
appears to be well described by a pure exponential (with slight 
variations between various magnetic boundary conditions), 
while the surface PDF is a stretched exponential only for 
some boundary conditions \citep[see][]{t2 TC00}.
It is also of interest to mention that the (plage)  
magnetoconvection simulations of \citet{t2 Vo03}
have been carried out by imposing an initial unipolar vertical magnetic 
field (which may lead to PDFs with a broad peak
at kG field strengths, if the imposed signed magnetic flux is sufficiently 
large), while the (internetwork)
simulations of \citet{t2 SN03}
were carried out with a horizontal magnetic flux advected in 
by entering fluid at the bottom of the computational box (which leads to 
stretched exponential PDFs).} Therefore, by choosing an exponential 
PDF instead of a Voigt PDF (or instead of any other better fitting 
function) we guarantee that we are not exaggerating our estimation 
of $\langle B \rangle$. For the more realistic case (b), we also 
found it important to compare the resulting  magnetic energy with 
that corresponding to the kG fields of the network patches 
\citep{t2 TB04}.

As shown in Fig.~2, for the standard case (a) of a single-value
microturbulent field we found that 
$\langle B \rangle\approx60$\,G leads
to a notable agreement with the observed $Q/I$. Note that 
the strength of the hidden field required to explain the $Q/I$ 
observations seems to decrease with height in the atmosphere, from 
the 70\,G needed to explain the observations at $\mu=0.6$ to 
the 50\,G required to fit the observations
at $\mu=0.1$. This corresponds to an approximate height range 
between 200 and 400\,km above the solar visible ``surface''.

Concerning the case of an exponential PDF, we see in Fig.~2 that 
$\langle B \rangle\approx130$\,G yields a fairly good fit to the observed 
fractional linear polarization.\footnote{Note in Fig.~2 that the 
$\langle B \rangle$ required to fit the $Q/I$ observations at 
$\mu=0.6$ is significantly larger than 130\,G, while that needed to 
fit the observations at $\mu=0.1$ is smaller.} In this more 
realistic case
$E_{\rm m}={\langle B^2 \rangle}/8{\pi}\approx 1300\,{\rm erg\,cm^{-3}}$ 
(i.e., $\langle B^2 \rangle^{1/2}\approx180$\,G), which is about 
20\% of the kinetic energy 
density produced by convective motions at a height of 200\,km in 
the 3D photospheric model. As pointed out by \citet{t2 TB04},
for this case the total magnetic energy stored in the internetwork 
regions turns out to be larger than that corresponding to the kG 
fields of the network patches. 

\begin{figure}[!t]
\centering
\includegraphics[width=10.0cm]{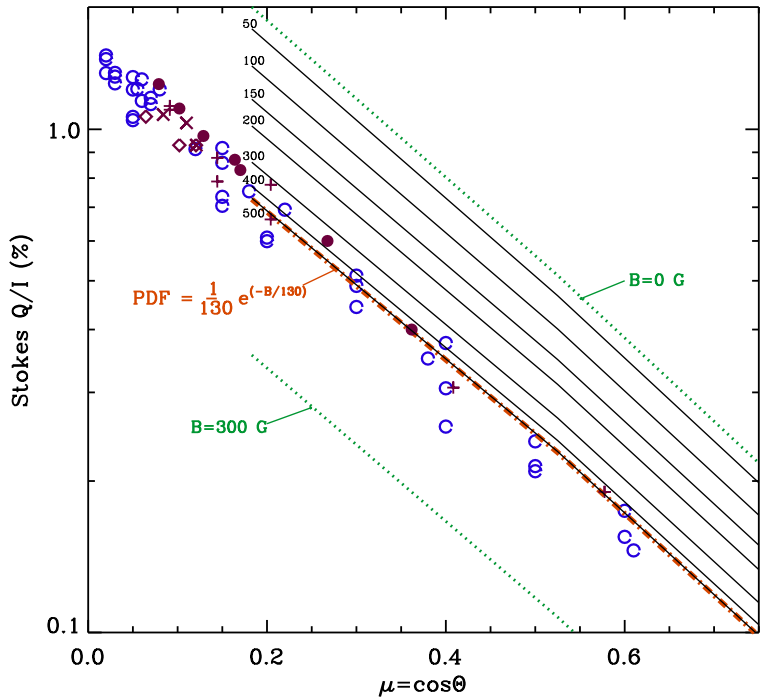}
\caption{This figure is similar to Fig.~2, with the black 
dashed-dotted line showing the same good overall fit obtained 
with an exponential PDF with $\langle B \rangle=130$\,G. 
The two dotted lines labeled $B=0$\,G and $B=300$\,G 
are also identical to those of Fig.~2.
In contrast, the solid-lines labeled by the values 
$B_{\rm final}=50,100,150,200,300,400,500$\,G 
show the calculated $Q/I$ amplitudes imposing a cut-off
of the exponential PDF at $B_{\rm final}$. This figure shows 
that, in order to be able to fit the $Q/I$ observations, we 
need to take into account all the magnetic fields with 
strengths between zero and 500\,G, whose filling 
factor is 98\%. The remaining magnetic fields of the exponential 
PDF above the cut-off of 500\,G correspond to a filling factor of 2\%,
and they make no significant contribution to the Hanle depolarization.}
\label{t2 fig:Fig3}
\end{figure}

\Citet{t2 DC06}
have recently arrived at similar conclusions by combining magnetic 
field strength measurements based on the Zeeman effect in the 
\ion{Fe}{i} lines at 6301.5\,\AA, 6302.5\,\AA, 15648\,\AA, 
and 15652\,\AA, and the Hanle effect in the \ion{Sr}{i} 4607\,\AA\ 
line. They approximated the total PDF with
$P(B)=w\,P_{\rm H}(B)+(1-w)\,P_{\rm Z}(B)$, where $P_{\rm Z}(B)$ is a
PDF inferred from the Zeeman signals, $P_{\rm H}(B)$ a PDF that fits 
the Hanle signals, and $w\approx1$. The fact that these authors
arrive at conclusions similar to those of \citet{t2 TB04}, can be
explained by noting that: 1) the fraction of quiet Sun producing the 
Zeeman polarization signals was only $\sim1.5\%$; 2) they used our 
value for the depolarization factor of the \ion{Sr}{i} 4607\,\AA\ line
(${\cal D}=0.4$); and 3) they imposed the reasonable constraint that 
the magnetic energy density has to be significantly smaller than 
the kinetic energy density of the granular motions. What makes their work different from ours 
is that their $P(B)$ shows a very significant peak at large 
field strengths (e.g., at 1750\,G for an atmospheric height $h=0$\,km,
and at 650\,G for $h=250$\,km), which leads them to conclude that kG 
fields in the internetwork regions of the quiet Sun still represent
a significant fraction of the total magnetic energy. This peak in their
PDF is a direct consequence of the peak that their $P_{\rm Z}(B)$ also
shows at kG field strengths, which they claim to be realistic, 
because they believe that the Zeeman polarization signals observed in the 
\ion{Fe}{i} lines at 6301.5\,\AA\ and 6302.5\,\AA\ indicate the 
presence of a significant amount of small-scale kG field concentrations 
within the internetwork regions \citep[see also][]{t2 DC03}.
Future investigations will clarify which is the true fraction of (internetwork) quiet Sun occupied by kG fields. In any case, it is important to point 
out that the small-scale magnetic fields that have been detected via 
the Zeeman effect (whose filling factor is of the order of only a few 
percent) make no significant contribution to the ``observed'' magnetic 
depolarization in the \ion{Sr}{i} 4607\,\AA\ line. In other words, 
the ``hidden'' magnetic field inferred by \citet{t2 TB04}
with the analysis of the Hanle effect carries a vast amount of unsigned 
flux and energy, independently of whether or not we have kG fields 
in the internetwork regions of the quiet Sun.

We must also mention that, in order to be able to fit the 
observed polarization amplitudes,
we need to take into account all the microturbulent magnetic fields with 
strengths between 0 and 500\,G (i.e., according to our exponential 
distribution of field strengths characterized by 
$\langle B \rangle=130$\,G), which altogether correspond to 
a filling factor of 98\% (see Fig.~3). 
The remaining magnetic fields of our 
exponential distribution, with strengths greater than 500\,G, have 
correspondingly a filling factor of only 2\%, and they do not 
contribute significantly to the ``observed'' Hanle depolarization. 
In conclusion, contrary to what
was argued by \citet{t2 SA05},
a detailed theoretical modeling 
of the scattering polarization of the \ion{Sr}{i} 4607\,\AA\ line does 
allow us to estimate the unsigned magnetic flux and magnetic energy 
existing in the quiet solar photosphere.

Taking into account that most of the solar
surface is occupied by the quiet internetwork regions
of mixed polarity fields, it is clear that the results
of our 3D analysis of the scattering polarization 
observed in the \ion{Sr}{i} 4607\,\AA\ line
might have far-reaching implications in solar and stellar physics.
The hot outer regions of the solar atmosphere
(chromosphere and corona) radiate and expand, 
which takes up energy. By far the largest energy
losses stem from chromospheric radiation, with a
total energy flux of 
$\sim10^7\,\rm erg\,cm^{-2}\,s^{-1}$ 
\citep{t2 AA89}.  
The magnetic energy density corresponding to the simplest
(and most conservative) model with $\langle B \rangle\approx60$\,G 
is 140\,erg\,cm$^{-3}$, which leads to an energy flux comparable
to the chromospheric energy losses, when using either
the typical value of $\sim1\,{\rm km\,s^{-1}}$
for the convective velocity, or the Alfv\'en speed, 
$v_{\rm A}=B/(4{\pi}{\rho})^{1/2}$, where $\rho$ is the gas density. 
In reality, as pointed out above, the true magnetic energy density 
that is stored in the quiet solar photosphere at any given time 
during the solar cycle is very much larger than 140\,erg\,cm$^{-3}$. 
For example, the magnetic energy density
corresponding to the (still conservative) case of an exponential
distribution of field strengths with $\langle B \rangle\approx130$\,G 
is 1300\,erg\,cm$^{-3}$, which implies an energy flux 10 times 
larger than the chromospheric radiative losses. Only a relatively 
small fraction would thus suffice to
balance the radiative losses of the solar outer atmosphere.

\begin{figure}[!t]
\centering
\includegraphics[width=8.2cm]{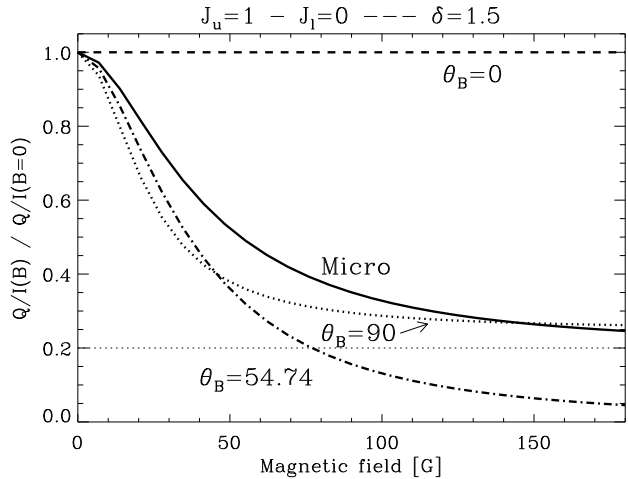}\hspace{5pt}
\caption{Variation of the Hanle depolarization factor of a 
triplet-type transition with the strength of the assumed 
volume-filling, single value magnetic field. While the solid 
line labelled ``micro'' 
corresponds to the isotropic microturbulent field case, the remaining 
curves correspond to cases with a fixed magnetic field inclination 
($\theta_B$), but with a random azimuth below the mean free path of 
the line photons.}
\label{t2 fig:Fig4}
\end{figure}

\section{\boldmath
Can the Inferred Depolarization be Explained by a Non-Isotropic 
Distribution of Small-Scale Fields with a Smaller $\langle B \rangle$?}

For various magnetic field topologies, Fig.~4 shows the depolarization 
factor corresponding to a triplet-type transition, like the 
\ion{Sr}{i} 4607\,\AA\ line. With the exception of the curve 
corresponding to the case of an isotropic microturbulent field that has
been considered so far, the remaining curves refer to cases with 
a fixed magnetic field inclination ($\theta_B$), but random azimuth 
at sub-resolution scales (which ensures that Stokes $U=0$). 
As we see, a depolarization factor ${\cal D}=0.4$ can be obtained 
with $\langle B \rangle\approx 45$\,G for random-azimuth cases 
with significantly inclined magnetic fields (e.g., with 
inclinations around the Van Vleck angle 
$\theta_{\rm VV}\approx 54\fdg74$), while it requires
$\langle B \rangle \approx 70$\,G for the isotropic-field case. 

With the exception of the $\theta_B=90^{\circ}$ case,
this type of random-azimuth fields would produce ubiquitous circular 
polarization signals when observed at disk center (unless we assume 
that we have pairs of random-azimuth magnetic field distributions 
of fixed inclination but with opposite orientations on 
sub-resolution scales). Obviously, our only aim with this illustration 
for non-isotropic distributions of micro-structured fields is to 
point out that, although smaller, the inferred mean magnetic field 
strength would still be very significant.

\section{The Hanle Effect in Ti\,{\sc i} Lines}

The previous sections have shown that the determination of the 
strength of the ``hidden'' field via the Hanle effect in the 
\ion{Sr}{i} 4607\,\AA\ line is model-dependent, in the sense that 
it relies on radiative-transfer calculations of the $Q/I$ 
amplitude of the strontium line in the absence of magnetic fields. 
We believe that our estimate is reliable because of the realistic 
3D model of the solar photosphere adopted. Nonetheless, it is 
important to find model-independent clues that magnetic 
depolarization by a tangled field is really at work in the quiet 
solar photosphere. 

As pointed out by \citet*{t2 MS04},
such ``direct'' evidence can be obtained by comparing the 
observed $Q/I$ amplitudes of the 13 lines of the  
$a^{5}F{-}y^{5}F^{0}$ multiplet of \ion{Ti}{i}, since
the 4536\,\AA\ line is completely insensitive to magnetic fields
(lower and upper levels with zero Land\'e factors), while the
remaining lines are sensitive to the Hanle effect. Interestingly, in
the atlas of \citet{t2 Ga02}, 
the 4536\,\AA\ line is the only one of that multiplet which shows a 
sizable scattering polarization signal. This fact was interpreted 
by \citet{t2 MS04} as a direct evidence of the existence of a ubiquitous,
tangled magnetic field at sub-resolution scales.

\begin{figure}[!t]
\centering
\includegraphics[width=8.1cm]{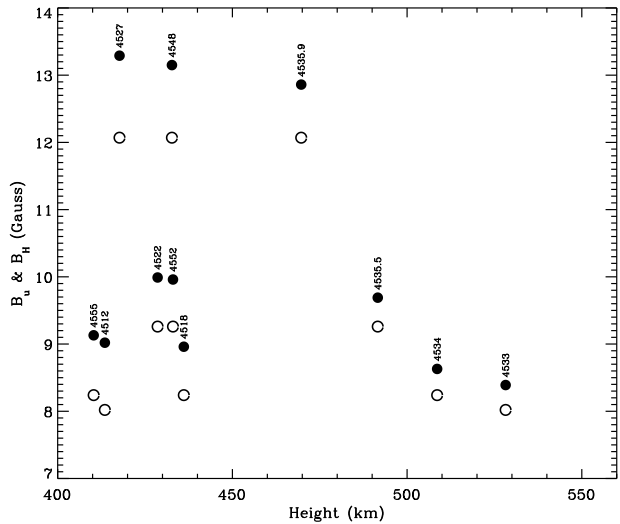}
\caption{Critical magnetic field strength for the upper-level 
Hanle effect in the lines of the $a^{5}F{-}y^{5}F^{0}$ multiplet 
of \ion{Ti}{i}, plotted against the height $H$ at which 
$\tau_\mathrm{line-core}=1$, for a simulated observation along 
$\mu=0.1$ in a semi-empirical model of the solar atmosphere. 
The lines at 4536\,\AA\ ($H\approx 450$\,km) and 4544\,\AA\ 
($H\approx 410$\,km) are not shown because they
have zero Land\'e factor for the upper level. ($\circ$) Values 
determined from Eq.~(1); ($\bullet$) values determined from Eq.~(4) 
(which takes into account the influence of collisional quenching 
using the collisional depolarizing rates of \citealt{t2 De05}).}
\label{t2 fig:Fig5}
\end{figure}

Obviously, the fact that those spectral lines belong to the 
same multiplet does not imply that the atmospheric height, $H$, 
where ${\tau}_{\rm line-core}=1$ for an observation at $\mu=0.1$ 
is the same for all the lines. In fact, as shown in Fig.~5, 
this height varies between 400 and 550\,km, approximately. 
However, the fact that $H=450$\,km for the 4536\,\AA\ line, while 
several of the remaining lines have larger values of $H$, 
indicates that the most plausible way of understanding the observed 
$Q/I$ amplitudes is indeed to invoke magnetic depolarization. 
The determination of the strength of the hidden field, however, 
requires detailed radiative transfer modeling \citep{t2 ST06}.

\section{The Hanle Effect in Molecular Lines} 

As it was reviewed in the previous section, 
the Hanle effect in the \ion{Sr}{i} 4607\,\AA\ line 
suggests the presence, in the quiet regions
of the solar photosphere, of a distribution of tangled
magnetic fields at sub-resolution scales
with a mean field strength of the order of 100\,G,
when no distinction is made between granular and inter-granular points. 
It would be of great interest to observe scattering polarization 
signals with high spatial and temporal resolution, in order to measure 
the horizontal fluctuations of the $Q/I$ amplitudes, with the aim of
determining any possible variation in the strength of the hidden 
magnetic field between granular and inter-granular regions. 
A promising observational strategy to achieve this goal is 
filter-polarimetry with a suitable tunable filter or with a 
Fabry-Perot interferometer. 
These types of observations are at present feasible, at least 
for the spectral lines which show the largest $Q/I$ amplitudes in 
the Second Solar Spectrum (that is, \ion{Sr}{i} 4607\,\AA, 
\ion{Ba}{ii} 4554\,\AA, and \ion{Ca}{i} 4227\,\AA),
but we will probably have to wait until the next solar 
polarization workshop to see the first results.

Is there any possibility of investigating, with a theoretical 
analysis of the observations of \citet{t2 Ga00,t2 Ga02},
whether or not the strength of the hidden field fluctuates on 
the spatial scales of the solar granulation pattern? At first sight 
this task might appear impossible, given that such observations of 
the Second Solar Spectrum 
lack spatial and/or temporal resolution. However, as shown by 
\citet{t2 TB04},
the joint analysis of the Hanle effect in the C$_2$ lines (via 
the differential Hanle effect technique described in
\citealt{t2 TB03a})
and in the \ion{Sr}{i} 4607\,\AA\ line (via the previously 
reviewed 3D modeling approach) actually allows this possibility.
This result and the intensive research work done to make this type 
of Hanle-effect diagnostic feasible are reviewed below.

\subsection{The Observational Discovery}

One of the interesting observational discoveries of the last decade
is that several diatomic molecules in the quiet regions of the 
solar photosphere, such as MgH and C$_2$, show conspicuous linear 
polarization signals when observing on-disk in quiet regions 
close to the solar limb \citep{t2 SK97}.
These observations of scattering polarization were carried out 
during a minimum of the solar activity cycle.

The observations of molecular scattering polarization by 
\citet{t2 SK97}
were confirmed by \citet{t2 Ga00},
and also by the full Stokes-vector observations of 
\citet{t2 TB01b},
which in addition showed that in the observed quiet-Sun regions 
$U/I\approx0$ and $V/I\approx0$. Both observing runs took place 
during the maximum phase of the solar activity cycle.

It is also of interest to mention that spectro-polarimetric
observations by \citet{t2 FA02}
seem to indicate that at least the C$_2$ lines show linear polarization 
signals also when observed just above the solar limb, 
where, as it was shown by \citet{t2 Pi68},
the intensity profiles of such molecular lines stand out in emission. 
It would be important to improve (and ideally confirm) such off-limb 
observations, because comparing and modeling on-disk vs.\ off-limb 
scattering polarization amplitudes is of great diagnostic interest 
\citep[see][]{t2 TB03a}. 

More recently, \citet{t2 Ga03} and \citet{t2 Sten03}
have pointed out that the CN violet band also shows very 
conspicuous scattering polarization signals, with a curious 
regular $Q/I$ pattern that was explained by \citet{t2 AT03}.

\subsection{Are Molecular Lines Immune to the Hanle Effect?} 

A comparison of Gandorfer's observations (taken between June 1999 
and May 2000) with those of \citet[][taken in April 1995]{t2 SK97}
indicated that the molecular scattering polarization amplitudes 
observed on-disk at a given distance from the solar limb 
show no variation with the solar cycle, in sharp contrast with 
the behavior of many (stronger) atomic lines 
\citep[e.g.,][]{t2 Ga00,t2 Sten03}.
This was
from the beginning regarded as an enigma without a
clear solution at hand.

\citet*{t2 Be02} 
claimed that the solution of this enigma was that the molecular lines 
that show measurable polarization amplitudes are simply ``immune'' to 
the Hanle effect because their Land\'e factors ($g$) are very small, compared
to the Land\'e factors of atomic lines that are typically $\sim1$. 
However, during the 3rd Solar Polarization Workshop (SPW3), 
\citet{t2 La03} and \citet{t2 TB03a}
pointed out that this was irrelevant to 
the problem, because one must also take 
into account that the radiative lifetimes ($t_{\rm life}$) of molecular 
levels are generally longer than those of atomic lines. As a result, 
the critical Hanle field defined in Eq.~(1) turns out to be similar 
for both atomic and molecular lines (e.g., $B_{\rm H}\approx23$\,G for 
\ion{Sr}{i} 4607\,\AA, compared to $B_{\rm H}\approx8$\,G for C$_2$ 5161.84\,\AA).

\subsection{\boldmath
The Differential Hanle Effect for C$_2$ Lines} 

\citet{t2 TB03a} reported on a very powerful tool to explore hidden 
magnetic fields at sub-resolution scales: a Hanle-effect line-ratio 
technique for the C$_2$ lines of the Swan system (currently known as 
the ``differential Hanle effect'' technique for C$_2$ lines). 
The basis of this technique becomes obvious if we look at the
plots of Fig.~6, which show the critical Hanle fields for the upper 
levels of the C$_2$ lines \citep[see also][]{t2 TB03c,t2 AR04}.
%
\begin{figure}[!t]
\centering
\includegraphics[height=3.1cm]{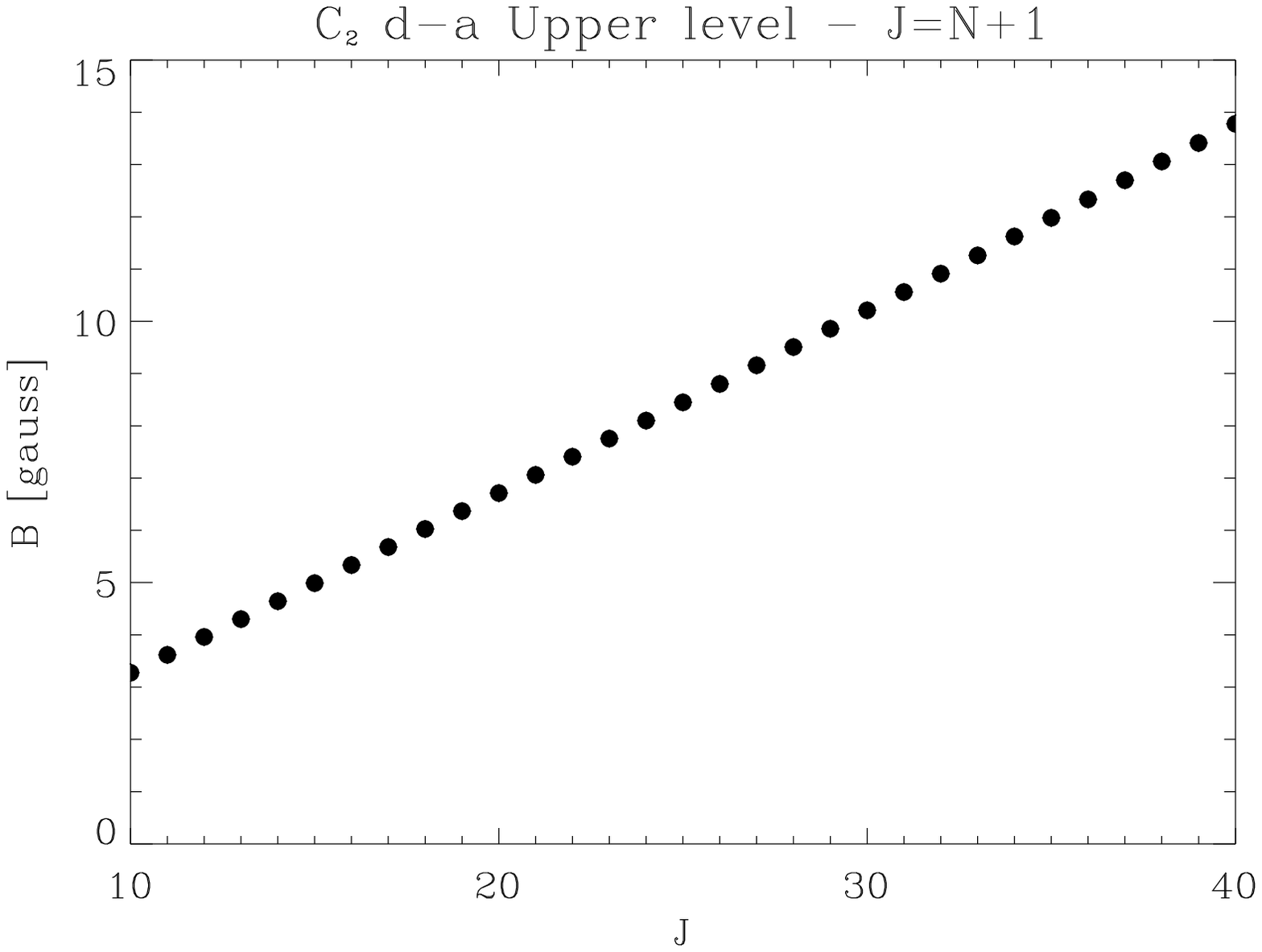}\hspace{5pt}
\includegraphics[height=3.1cm]{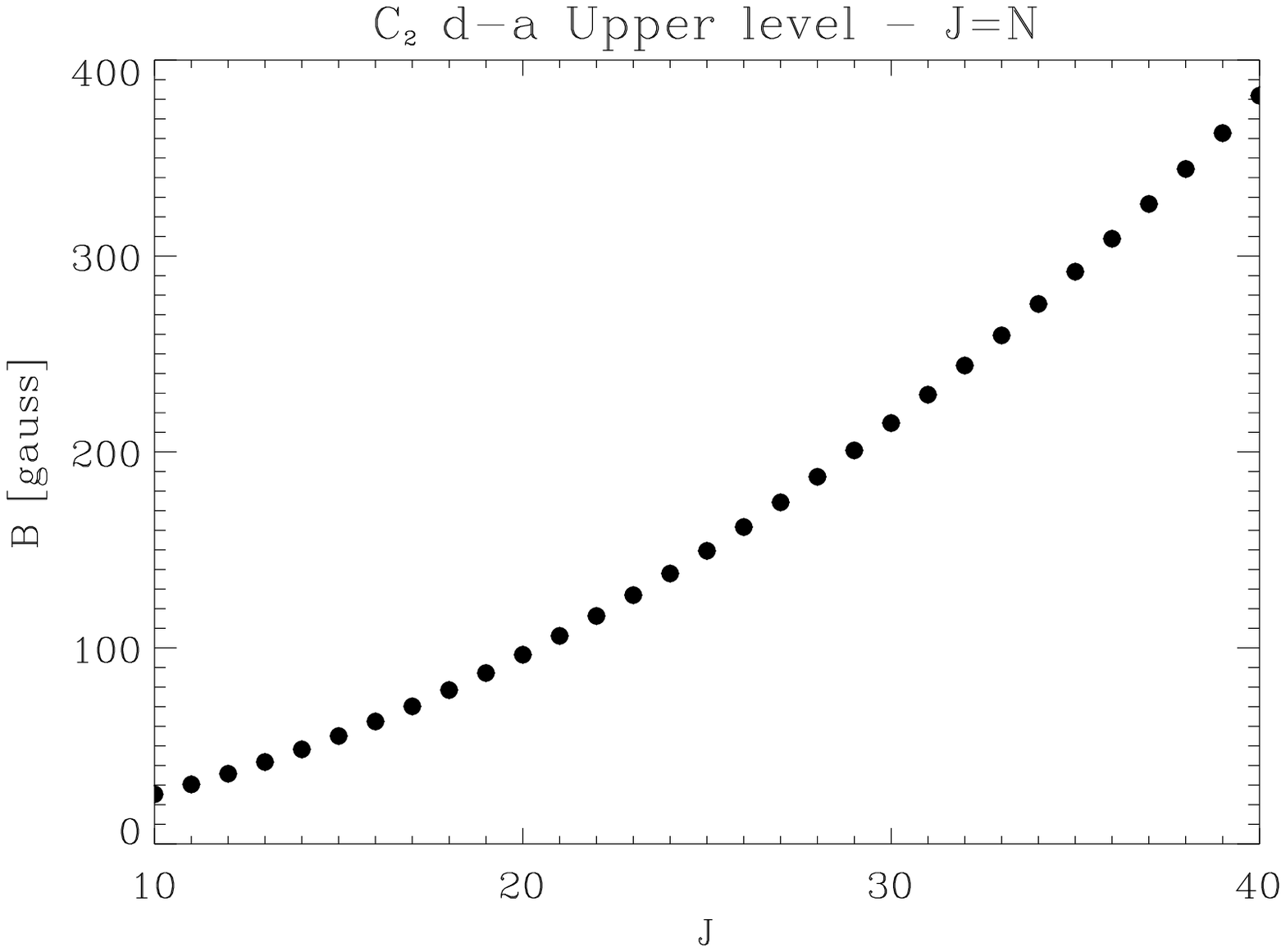}\hspace{5pt}
\includegraphics[height=3.1cm]{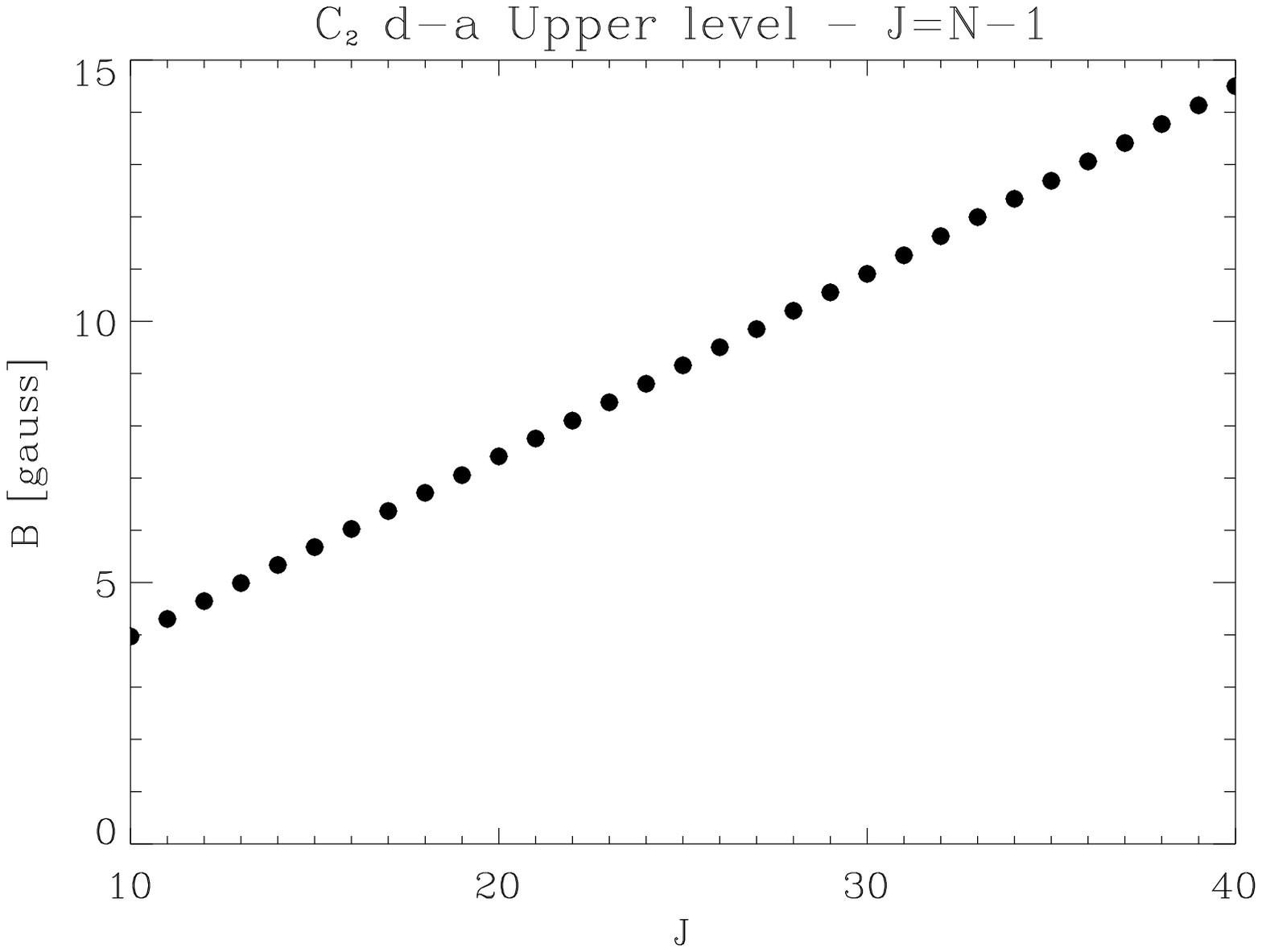}
\caption{Critical Hanle fields for the upper levels of the C$_2$ lines 
of the Swan system. Note that the critical Hanle fields are much 
larger for the levels with $J=N$ (center) than for those with 
$J=N+1$ (left) or $J=N-1$ (right).}
\label{t2 fig:Fig6}
\end{figure}
\begin{figure}[!t]
\centering
\includegraphics[width=6.1cm]{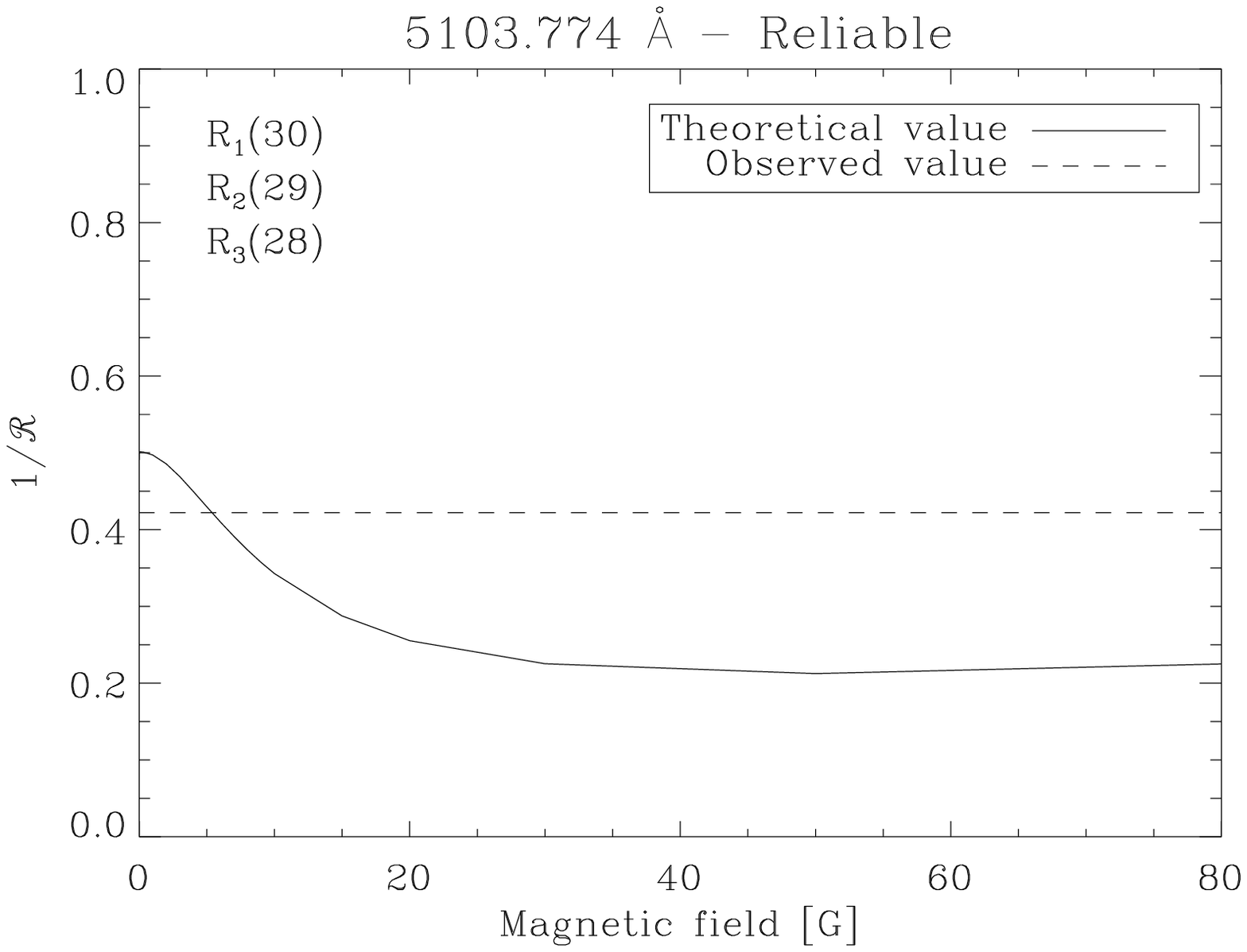}\hspace{5pt}
\includegraphics[width=6.1cm]{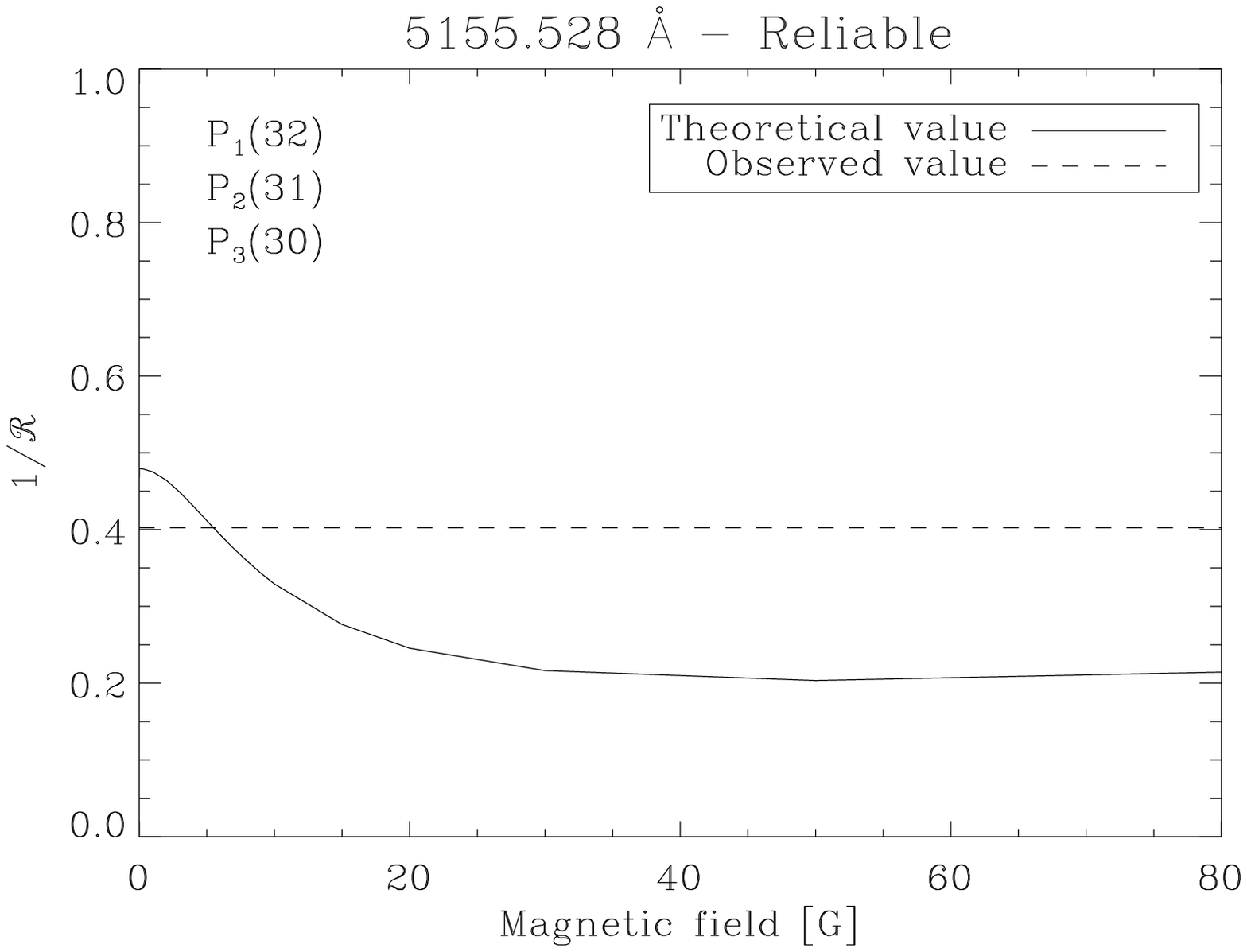}\vspace{5pt}
\includegraphics[width=6.1cm]{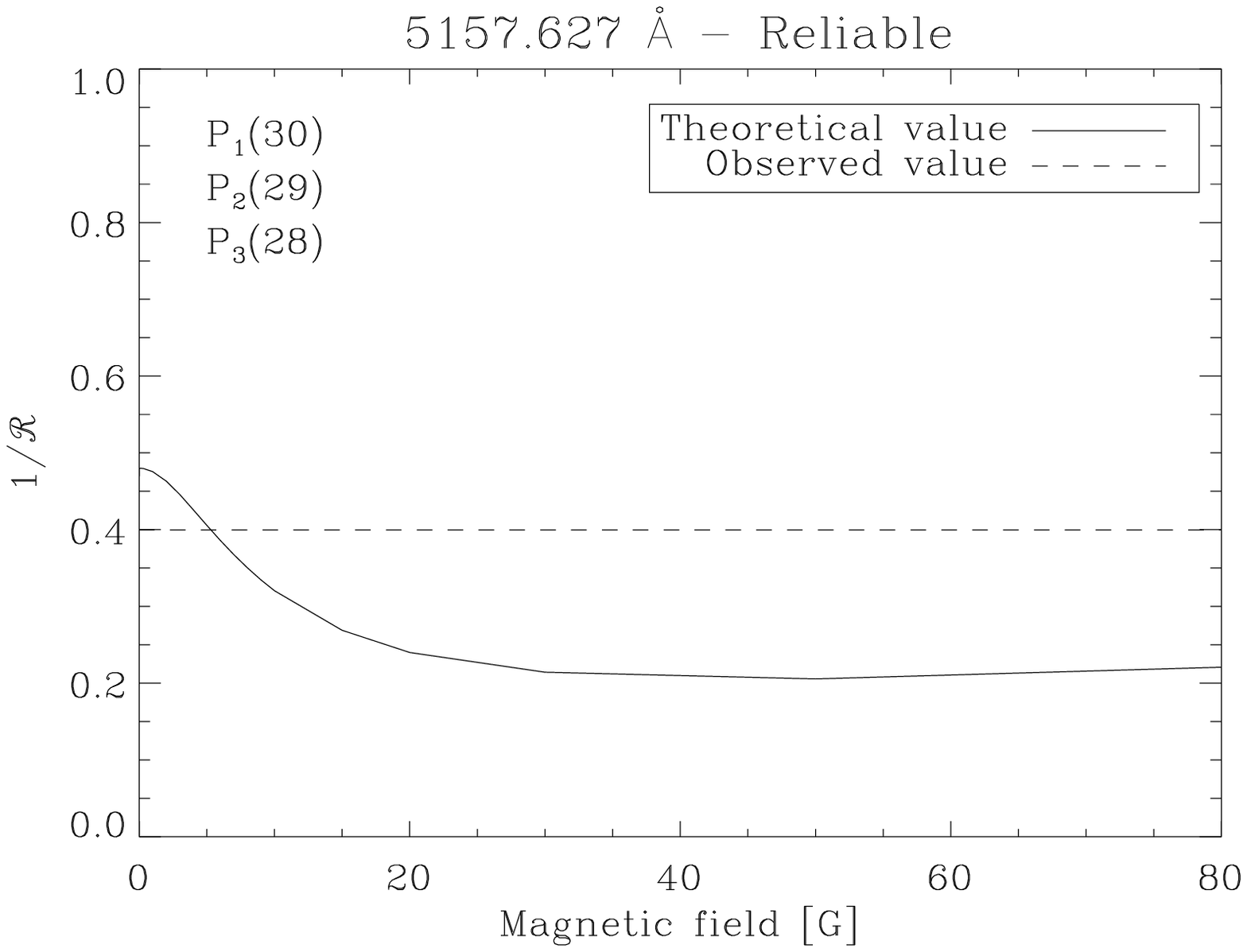}\hspace{5pt}
\includegraphics[width=6.1cm]{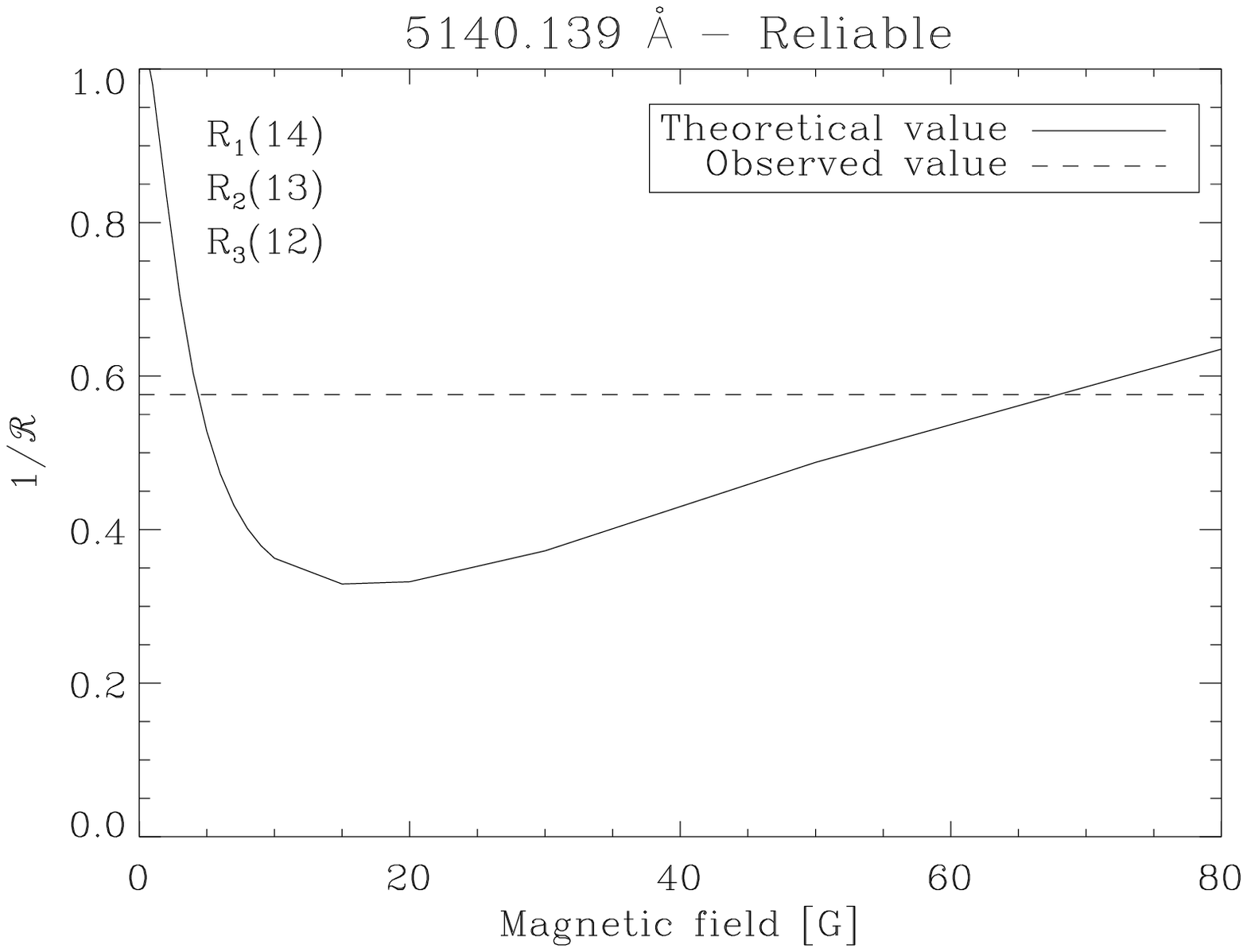}
\caption{Examples of the observed quantity $1/{\cal R}$ (dotted line) 
versus the calculated one (solid line) for increasing values of 
the strength of the assumed single value microturbulent field. 
(The quantity ${\cal R}$ is defined at the end of Section 6.3.)
The first three panels show examples of our blended C$_2$ lines with 
$J>20$, while the fourth panel corresponds to the unblended 
R-triplet mentioned in the text. Note that, in principle, the 
observed polarization ratio in this unblended triplet can be 
explained by two rather different strengths of the 
assumed microturbulent field. Fortunately, the observed polarization 
ratio of the blended C$_2$ lines is only compatible with the 
weak-field solution, because the corresponding R$_2$(P$_2$) line can 
be considered as a suitable reference line.}
\label{t2 fig:Fig7}
\end{figure}

The critical Hanle fields corresponding to the upper levels of the 
R$_2$ (P$_2$) lines (see central panel of Fig.~6, for the levels with 
$J=N$) are much larger than those corresponding to the upper levels of 
the R$_1$ (P$_1$) and R$_3$ (P$_3$) lines (see left and right panels of
Fig.~6, for the levels with $J=N+1$ and $J=N-1$, respectively). This 
is the case because, while the lifetimes of all such $J$-levels are 
similar, their Land\'e factors are much smaller for the levels with
$J=N$ than for those with $J=N\pm1$. As a result, the magnetic field 
strength needed for a significant Hanle-effect depolarization
is considerably smaller for the R$_1$ (P$_1$) and R$_3$ (P$_3$) lines 
than for the R$_2$ (P$_2$) lines. For instance, 
$B_H\approx8$\,G for the P$_1(J=25)$ line at 5161.67\,\AA\ 
and for the P$_3(J=23)$ line at 5161.84\,\AA, while $B_H\approx140$\,G 
for the P$_2(J=24)$ line at 5161.73\,\AA . Therefore, if we select 
R$_2$ and/or P$_2$ lines of C$_2$ having a ``sufficiently large"
critical Hanle field --i.e., significantly larger than the mean field 
strength of the internetwork regions in the photosphere-- it should 
be possible to use them as ``reference" lines to infer the strength of 
the hidden field via the application of the differential Hanle effect.
This is done through a direct comparison between the linear 
polarization amplitude observed in the R$_2(J)$ (P$_2(J)$) line selected
(whose $Q/I$ value is assumed to correspond to the non-magnetic 
reference case) and the $Q/I$ observed in the  R$_3(J-1)$ (P$_3(J-1)$) line. 

But which P$_2$ and/or R$_2$ lines have a ``sufficiently large'' 
critical Hanle field? 
According to our Hanle-effect analysis of the 
\ion{Sr}{i} 4607\,\AA\ line, the mean field strength in the photospheric
internetwork regions is $\langle B \rangle\sim100$\,G. 
As seen in the central panel of Fig.~6, only the R$_2$ (P$_2$) lines 
with $J>20$ have a $B_{\rm H}>100$\,G. For this reason, the first 
important step is to choose suitable
pairs of C$_2$ lines having $J>20$. 
For each pair, we selected lines that
differ in their sensitivity to the Hanle effect, 
namely P$_2(J)$ and P$_3(J-1)$ lines, or R$_2(J)$ 
and R$_3(J-1)$ lines, but in either case with $J>20$. The fact 
that such R$_2(J)$ (P$_2(J))$ lines are blended with the 
R$_1(J+1)$ (P$_1(J+1)$) lines 
implies that 
{\em in the absence of magnetic fields} the ratio, ${\cal R}$, of 
the observed fractional 
linear polarizations between each particular R$_2$ (P$_2$) line and 
the corresponding R$_3$ (P$_3$) line is not unity, but a number 
between 1 and 2. For example, for the case of a perfect overlap 
between such two lines we have ${\cal R}=2$ in the absence of 
magnetic fields, while a ratio ${\cal R}$ notably larger than 2 
would indicate the presence of a substantial 
Hanle depolarization \citep{t2 TB03a}.


\begin{figure}[!t]
\includegraphics[width=6.5cm]{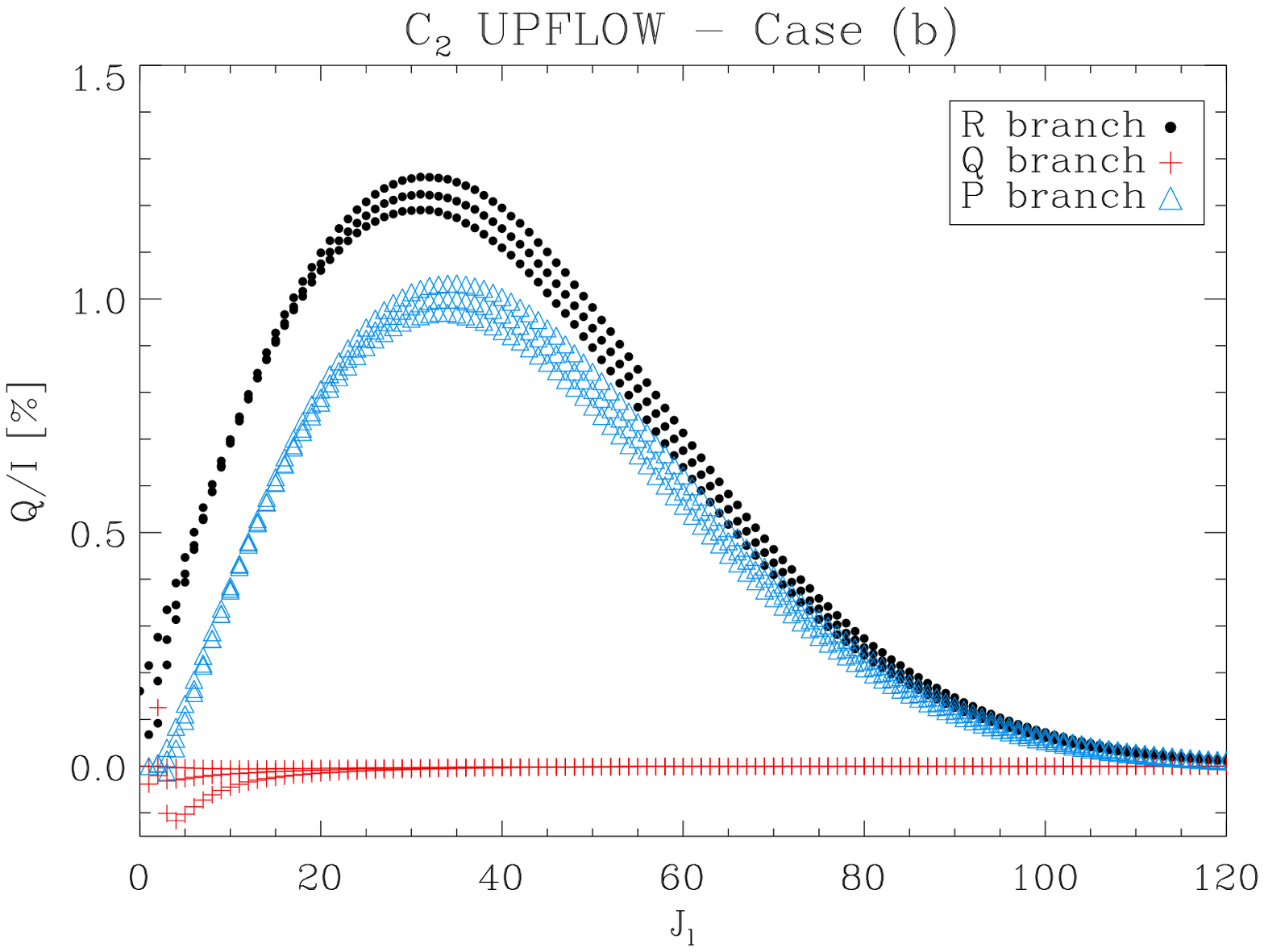}\hspace{10pt}%
\includegraphics[width=6.5cm]{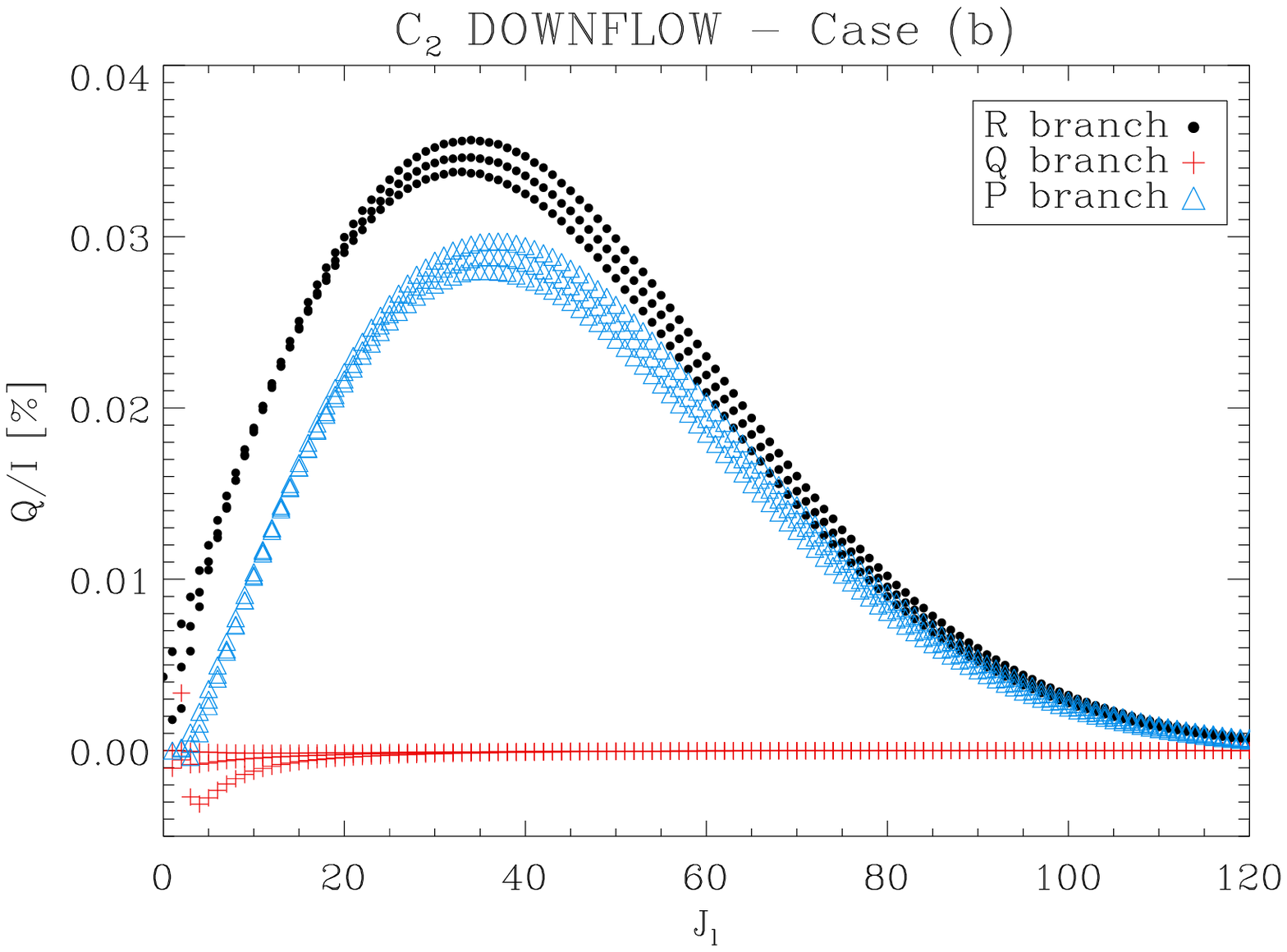}\vspace{5pt}
\includegraphics[width=6.5cm]{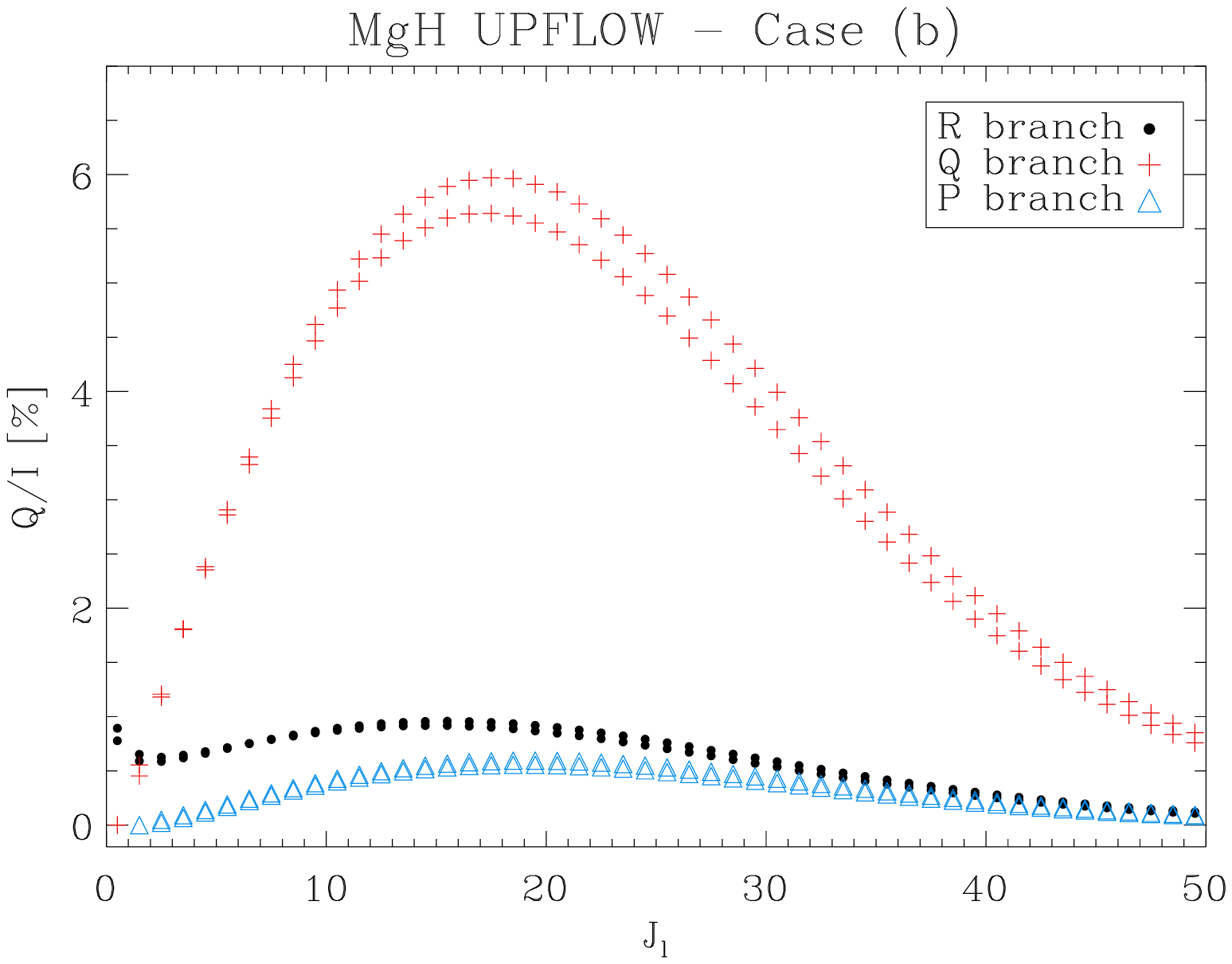}\hspace{10pt}%
\includegraphics[width=6.5cm]{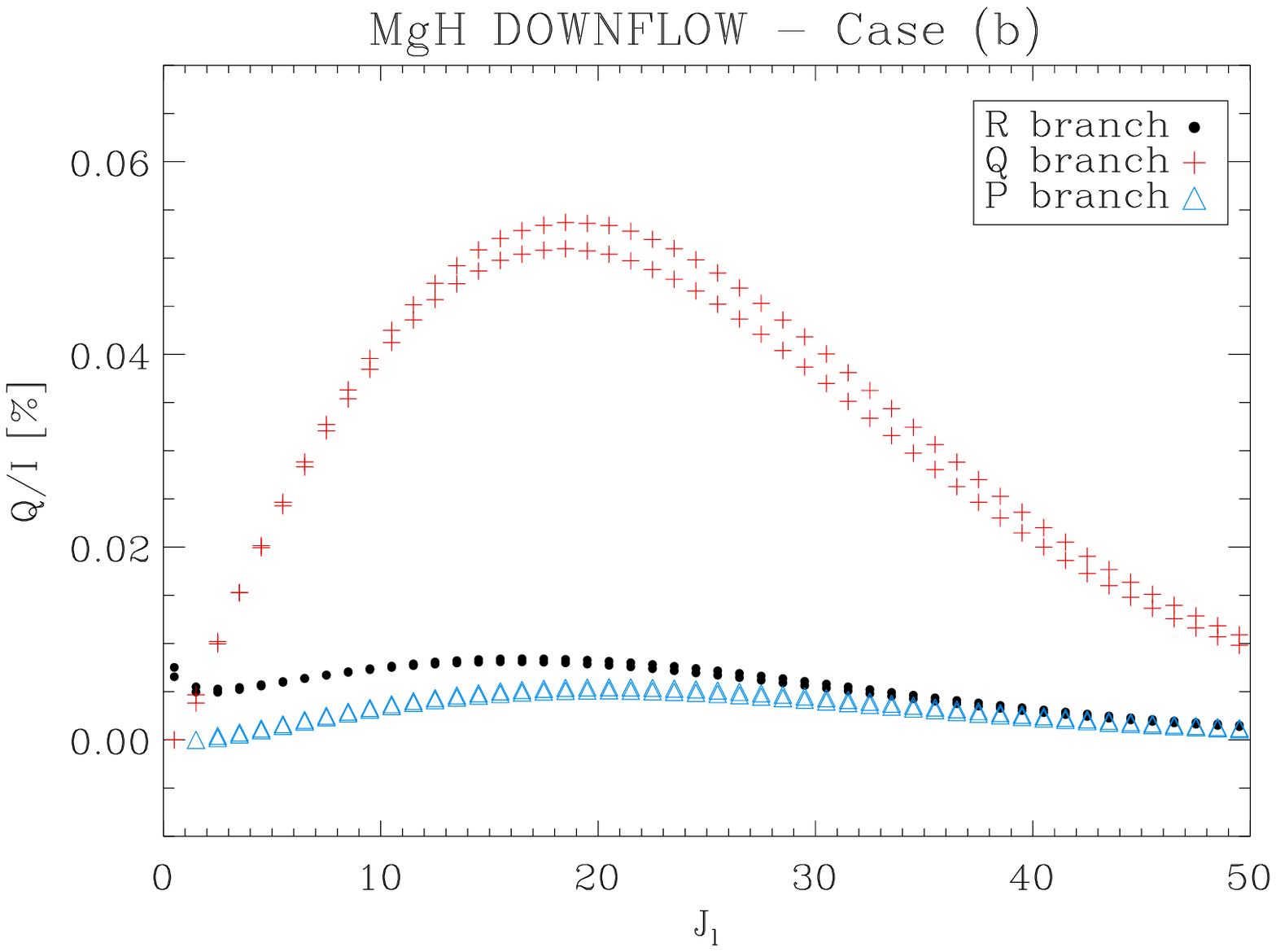}
\caption{This figure illustrates that the observed scattering 
polarization amplitudes in the P and R lines of the Swan system of 
C$_2$, and in the $Q$ lines of MgH, are coming mainly from the 
(granular) upflowing regions of the quiet solar photosphere. 
Although the height we chose for this illustration is 250\,km, 
the height where $\tau_{\rm line-core}=1$ in a realistic model 
of the solar photosphere is smaller for the MgH lines than for 
the C$_2$ lines.}
\label{t2 fig:Fig8}
\end{figure}

\subsection{\boldmath
Applications of the Hanle-effect line-ratio technique for C$_2$ lines}

Our analysis of the observations by \citet{t2 Ga00}
of the scattering polarization signals of the C$_2$ lines with $J>20$ 
revealed very small depolarizations (see examples in the first 
three panels of Fig.~7). This led us to the conclusion that most of 
the volume of the photospheric regions that contribute to the 
observed molecular scattering polarization is occupied by very 
weak fields \citep{t2 TB03a}.
In fact, during his talk at the SPW3, Trujillo Bueno already reported 
that the application of 
the Hanle-effect line-ratio technique to the C$_2$ lines suggests 
that $\langle B \rangle\approx10$\,G. A refined analysis 
leads us to
conclude that $\langle B \rangle\approx15$\,G if we assume an 
exponential PDF, or $\langle B \rangle\approx7$\,G for the simpler case of a single value microturbulent field \citep{t2 TB04}.
So this technique proves itself a very powerful diagnostic tool for the
investigation of ``turbulent" magnetic fields, as it has been confirmed 
by other authors, who soon after SPW3 embraced the same technique
to apply it to their investigations. For example, \citet{t2 BF04}
applied it to three unblended C$_2$ lines that are thought to 
be more sensitive to the weaker fields (namely, the R$_1(J=14)$ line 
with $B_{\rm H}\approx4$\,G, the R$_2(J=13)$ line with 
$B_{\rm H}\approx40$\,G and the R$_3(J=12)$ line with 
$B_{\rm H}\approx4$\,G). Note that using R$_2(J=13)$ as a reference
line is justified only if the mean field strength of the photospheric regions 
where the observed molecular scattering polarization is mainly produced is 
{\em significantly} smaller than 40\,G. 
Fortunately, this is precisely the case \citep[see Sect.~6.5, and][]{t2 TB03a}.

The R-triplet used by \citet{t2 BF04} is useful because
the R$_2$ reference line is not blended with the R$_1$ line, which 
makes it easier to detect any possible magnetic depolarization (see 
the fourth panel of Fig.~7). They reported 
$\langle B \rangle\approx15$\,G for the case of a single value 
microturbulent field, as also did \citet{t2 FA03}
via a complicated analysis of some of the C$_2$ lines with $J>20$ that 
we had previously considered. 
The disagreement with the value of 7\,G determined by 
\citet{t2 TB04} for the same model can be explained when we consider that
both \citet{t2 BF04} and \citet{t2 FA03}
overestimated the Einstein coefficients by a factor 2, 
overlooking the fact that 
$\Lambda$-doubling is not present in C$_2$
\citep[see, e.g.,][and \citealt{t2 La06}]{t2 He50}. 

\subsection{Which Regions Produce the Observed Molecular 
Scattering Polarization?}

In the previous section, we showed how the
analysis of the Hanle effect in the ${\rm C}_2$
lines of the Swan system also supports the hypothesis 
of a hidden magnetic field at sub-resolution scales. However
it suggests a mean field strength $\langle B \rangle\sim10$\,G, 
which is much smaller than what is needed to explain the observations of 
the \ion{Sr}{i} 4607\,\AA\ line presented earlier, which point instead 
to a hidden field with $\langle B \rangle\sim100$\,G. A resolution 
of this ``enigma'' was found when it was pointed out by \citet{t2 TB03a}
that the observed scattering polarization in very weak spectral lines, 
such as those of C$_2$ and MgH, is coming {\em mainly} from the upflowing 
regions of the quiet solar photosphere 
\citep[see also Fig.~2 of][]{t2 TB04}. 
One way to illustrate this result is by using the single-scattering approximation formula
\begin{equation}
\frac{Q}{I}\approx\frac{3}{2\sqrt{2}}\,
	\frac{\eta_I^{l}}{\eta_I^{l}+\eta_I^{c}}\,
	\frac{S_l}{B_{\nu}}\,{\cal A}{\cal W}_{2}\;,
\end{equation}
where most of the symbols have their usual meaning 
(e.g., ${\cal A}=J^2_0/J^0_0$ is the degree of anisotropy of the 
radiation field; cf.\ \citealt{t2 TB03a}), while 
${\cal W}_{2}$ is the generalized polarizability factor defined 
by \citet{t2 La03} (see also \citealt{t2 La06b}),
which is the correct one for the astrophysical case 
(see Fig.~5 of \citealt{t2 AT03},
for a comparison with the laboratory-case ${\cal W}_{2}$ used by 
\citealt{t2 Be02}).
Figure~8 shows the $Q/I$ values we obtained by applying this formula 
with the physical conditions encountered at a height of 250\,km in 
an upflowing cell center, and in a nearby downflowing intergranular 
lane, of a snapshot of the 3D hydrodynamical simulations of solar surface 
convection by \citet{t2 As00}.

In conclusion, the analysis of the scattering polarization observed 
without spatio-temporal resolution in C$_2$ and MgH lines is giving 
us empirical information on the distribution of hidden magnetic 
fields {\em mainly} in the (granular) upflowing regions of the quiet 
solar photosphere. 


\subsection{Evidence for a Fluctuating ``Hidden'' Field}

First of all, it is important to point out that 
the calculated scattering polarization amplitude of the 
(moderately strong) \ion{Sr}{i} 4607\,\AA\ line, for the 
zero-field reference case, gets significant contributions from 
both the upflowing and downflowing regions of the quiet 
solar photosphere.
In contrast, we just showed that the scattering polarization of 
weak molecular lines, like C$_2$, is mainly produced in the 
(granular) upflowing regions, and we concluded from our analysis
that these regions must be weakly magnetized, with
$\langle B \rangle\sim10$\,G. 

We now must ask how large the strength of the hidden field in the 
(intergranular) downflowing regions has to be, in order to be able 
to explain the inferred depolarization in the \ion{Sr}{i} 
4607\,\AA\ line. Interestingly, we find that the distribution  
of magnetic field strengths in the (intergranular) downflowing 
regions of the quiet solar photosphere must produce saturation for 
the Hanle effect in the \ion{Sr}{i} 4607\,\AA\ line formed 
there.\footnote{As seen in Fig.~2, saturation for the simplest case 
of a single value field occurs for $B\gtrsim 200$\,G.} For this reason, 
\citet{t2 TB04} concluded that the joint analysis of the Hanle 
effect in C$_2$ lines ($\langle B \rangle\sim10$\,G) and in the 
\ion{Sr}{i} 4607\,\AA\ line ($\langle B \rangle\sim100$\,G) 
suggests that the strength of the 
hidden magnetic field ``fluctuates'' on the spatial scales of the 
solar granulation pattern, with much stronger fields above the 
intergranular regions. 

The simplest intergranular PDF that produces Hanle-effect saturation 
for the \ion{Sr}{i} 4607\,\AA\ line in the intergranular regions
is that corresponding to a single value microturbulent field with 
$B\gtrsim 200$\,G filling the entire downflowing volume. A more realistic 
scenario results when taking into account that the true intergranular 
PDF should have a tail towards the kG field strengths, as suggested 
by numerical experiments on turbulent dynamos and/or magnetoconvection.  
A merely illustrative example of an intergranular PDF that satisfies 
such requirements is the Maxwellian PDF used by \cite{t2 TB04}
for the downflowing regions of the quiet solar photosphere, 
${\rm PDF}_{\rm downflows}=2.38\times10^{-8}\,B^2\exp(-B/456)^2$, 
which we obtained by fitting with a Maxwellian the strong-field part 
of the intergranular histogram of the Zeeman splittings observed by 
\citet{t2 Kh03} 
in the \ion{Fe}{i} lines at 1.56\,\micron. 
Note that this illustrative PDF implies that the filling factor of 
kG fields is $\sim2\%$ of the downflowing volume---that is, it 
does not exclude the possibility of some kG field concentrations in 
the internetwork regions \citep{t2 DC03}. We point out that with 
such a Maxwellian PDF for the {\em downflowing plasma}, most of 
the magnetic energy is carried by chaotic fields with
strengths between the equipartition field values and $\sim1$\,kG. A 
similar conclusion is reached with other plausible choices for the 
intergranular PDF (e.g., a Gaussian), in so far that they satisfy 
the following constraints:
\begin{enumerate}
\item[a)] to produce saturation of the Hanle effect in the 
\ion{Sr}{i} 4607\,\AA\ line;
\item[b)] to produce a Zeeman broadening of the intensity profiles of
carefully-selected near-IR lines that is compatible with the 
observations.\footnote{In this respect, 
we must point out that the Maxwellian PDF introduced earlier
has its maximum at a field strength that seems too high (456\,G), 
since it produces an excess of Zeeman broadening in the 
red wing of the 15648.5\,\AA\ line of \ion{Fe}{i}.}
\end{enumerate}
Obviously, with the Hanle effect we cannot distinguish between 
two magnetic field strengths having $B>B_{\rm satur}$ (where 
$B_{\rm satur}\approx200$\,G for the \ion{Sr}{i} 4607\,\AA\ line). 
For this reason, it is important to emphasize that a suitable way 
of constraining the parameters of the functional form chosen for 
the intergranular PDF is via a careful analysis of the Zeeman 
broadening of the {\em intensity profiles} of some carefully-selected 
near-IR lines that we have observed with the Tenerife Infrared Polarimeter. 

\subsection{\boldmath
Is Collisional Depolarization Significant for MgH and C$_2$?}

We already showed how the analysis of the Hanle effect in the C$_2$ lines 
of the Swan system can be carried out by applying the 
Hanle-effect line-ratio technique for C$_2$ lines, which leads to 
the conclusion that $\langle B \rangle \sim 10$\,G in the (granular) 
upflowing regions of the quiet solar photosphere 
\citep{t2 TB03a,t2 TB04}.
It is certainly fortunate that such technique is available, 
otherwise one would need a rather complex modeling of 
{\em multilevel} scattering polarization, given the fact that 
the two-level approximation is unsuitable for the C$_2$ lines 
\citep{t2 La03,t2 AT03}.

\begin{figure}[!t]
\centering
\includegraphics[width=9.0cm]{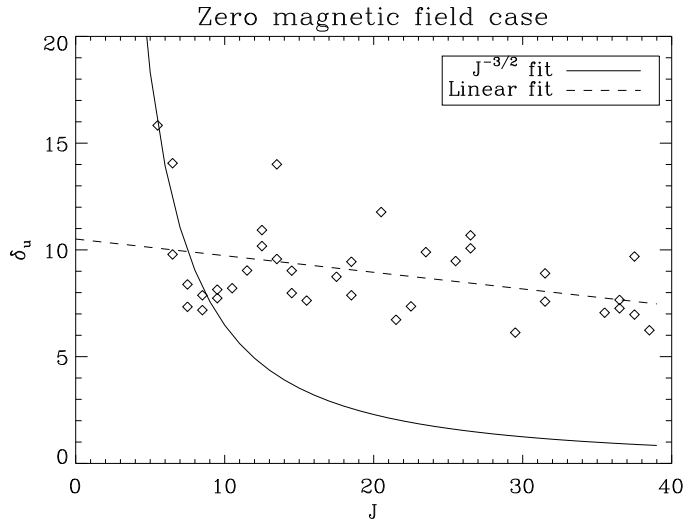}
\caption{For the case dominated by collisional depolarization 
mentioned in the text, this figure shows the collisional depolarizing 
rate that is needed to fit the scattering polarization amplitude 
of each of the unblended MgH lines observed by \citet{t2 Ga00},
for which $A_{ul}\approx10^{7}\,\rm s^{-1}$. The fact that the
${\delta}_u$ values required for the best fit are almost 
insensitive to $J$, instead of being proportional to $J^{-1.5}$ 
as predicted by the theoretical work of \citet{t2 De06},
suggests that the collisional depolarization inferred by \citet{t2 AT05}
is mainly caused by collisional transitions between different 
$J$-levels pertaining to the same vibrational and electronic state. 
While these types of collisional transitions are allowed for MgH, 
they are however strictly forbidden for C$_2$, which reinforces our 
conclusion that $\langle B \rangle \sim 10$\,G in the upflowing regions 
of the quiet solar photosphere.}
\label{t2 fig:Fig9}
\end{figure}

Which microturbulent mean field strength do we find via analysis 
of the observed scattering polarization signals in MgH lines?
To answer this question we need a detailed 3D modeling, in order 
to be able to estimate {\em correctly} the amplitude of the
zero-field linear polarization for each MgH line, which is produced 
by radiation scattering in the inhomogeneous solar 
photosphere.\footnote{Note that the scattering polarization in 
MgH lines comes mainly from the (granular) upflowing 
regions, where the vertical temperature gradients and the anisotropy 
factor are larger than in a 1D semi-empirical model.}
Fortunately, this 3D modeling of scattering polarization is
feasible, since a two-level molecular model for each particular 
MgH line provides a reliable approximation. In fact, 
\citet{t2 AT05}
have recently solved the 3D scattering polarization problem in each 
of the 37 unblended MgH lines that show significant $Q/I$ amplitudes 
in the Second Solar Spectrum atlas of \citet{t2 Ga00}.
Their radiative transfer approach is similar to that pursued by 
\citet{t2 TB04}
for the \ion{Sr}{i} 4607\,\AA\ line, adopting a
realistic 3D model of the solar photosphere as it results from the 
hydrodynamic simulations of solar surface convection by 
\citet{t2 As00}.
The aim was to investigate which combinations of field strengths 
and collisional depolarizing rates produce polarization amplitudes 
in agreement with the MgH observations of \citet{t2 Ga00}.
\citet{t2 AT05}
found that there are two possible regimes that 
lead to comparable best fits to the observed scattering 
polarization amplitudes:
\begin{enumerate}
\item[1)] a case dominated by collisional depolarization, but with the
possibility of a weak microturbulent field whose strength cannot be 
much larger than 10 or 20\,G (concerning the case of a single value
field), in order to preserve the goodness of the best fit;\footnote{We 
point out that $\langle B \rangle\lesssim10$\,G 
is obtained when the Einstein coefficients given by \citet{t2 We03}
are used, while assuming that Kurucz's $A_{ul}$-values are more 
accurate (as claimed by Bommier; private communication) then the 
approximate \textit{upper limit} would be 20\,G (which is still a factor 3 
smaller than the mean field strength inferred via 3D modeling of 
the \ion{Sr}{i} 4607\,\AA\ line polarization).}
\item[2)] a strongly magnetized case characterized by a microturbulent 
field with strength greater than a few $10^2$\,G, which 
nevertheless requires a small amount of collisional depolarization
in order to attain the best fit.
\end{enumerate}
Given that the observed scattering polarization comes mainly from the 
(granular) upflowing regions of the quiet solar photosphere, and that 
it is unrealistic to think that such regions are filled by a tangled 
magnetic field stronger than 100\,G, \citet{t2 AT05}
opted for the first case dominated by collisional depolarization. 
Interestingly, this case is also characterized by 
relatively weak fields (see Note 10), so that it is compatible with 
the result of $\langle B \rangle \sim 10$\,G
that was 
obtained directly via the application of the Hanle-effect 
line-ratio technique for the C$_2$ lines of the Swan 
system. That result, however, was based on the assumption that 
collisional depolarization is unimportant for the upper levels of 
C$_2$ lines. If this were not the case, then the inferred mean field 
strength would be larger, given that in reality the magnetic field 
strength required to produce a sizable Hanle depolarization is 
approximately given by \citep[e.g.,][]{t2 TB03a}
%
\begin{equation}
B\approx(1+{\delta_u})\,B_{\rm H}\, ,
\end{equation}
where $B_{\rm H}$ is given by Eq.~(1), and 
${\delta}_u=D^{(2)}\,t_{\rm life}\approx D^{(2)}/A_{ul}$. 
Therefore, it seems that, while collisional depolarization is 
essential for the MgH lines, in order to retrieve $\langle B
\rangle \sim 10$\,G, for the same reason it must be insignificant 
for the C$_2$ lines.
We think that this is actually the case, given that C$_2$ is a 
homonuclear molecule in which collisional transitions between levels 
$J$ and $J\pm1$ are strictly forbidden (see pages 131 and 238 of 
\citealt{t2 He50}). In fact, as suggested by Fig.~9, this type of 
collisional transitions between nearby but different $J$-levels
pertaining to the same vibrational and electronic state are probably 
the main cause of any collisional depolarization possibly
occurring in the other molecular lines of the Second Solar Spectrum.

\section{Conclusions and Outlook for the Future}

A joint analysis of the Hanle effect in molecular lines 
and in the \ion{Sr}{i} 4607\,\AA\ line leads to the conclusion that there
is a vast amount of ``hidden'' magnetic energy and (unsigned) magnetic
flux in the internetwork regions of the quiet solar photosphere, carried
mainly by rather chaotic fields in the (intergranular) downflowing
plasma with strengths $B>B_{\rm satur}\approx200$\,G. This
result, complemented with the (upper-limit) constraint imposed by our
analysis of the Zeeman broadening of the intensity profiles of carefully
selected near-IR lines, suggests that most of this energy is actually
carried by tangled magnetic fields at sub-resolution scales with
strengths between the equipartition field values and $\sim1$\,kG. 
This hidden magnetic energy in the internetwork regions of the quiet 
Sun is larger than that corresponding to the kG fields of the 
supergranulation network, and is more than sufficient to compensate 
the radiative energy losses of the solar outer atmosphere 
\citep{t2 TB04}.

Although the hidden magnetic field that we have discussed here 
carries most of the (unsigned) magnetic flux of the solar 
photosphere, we also must take into account the small-scale 
magnetic fields, with a filling factor of $\sim1$\%, that have been 
detected in the internetwork regions via the polarization induced 
by the Zeeman effect in \ion{Fe}{i} lines \citep[e.g.][]{t2 DC03,t2
Kh03,t2 SA03b}.
Even though such magnetic fields make no significant contribution 
to the ``observed'' Hanle depolarization (because their filling 
factor is so small), they might still carry a significant fraction 
of the total magnetic energy, mainly if the observed circular 
polarization signals in the 6301.5\,\AA\ and 6302.5\,\AA\ lines of 
\ion{Fe}{i} are really tracing kG fields concentrations, as claimed by 
\citet{t2 DC06}.
Given that kG field concentrations in the internetwork regions of 
the quiet Sun would have a more obvious impact on the magnetic 
coupling with the outer atmosphere \citep[e.g.,][]{t2 ScT03,t2 Wo06},
the determination of the true fraction of (internetwork) quiet 
Sun, occupied by magnetic fields with $B>1$\,kG, should be a high-priority
goal. In our opinion, a suitable strategy to achieve this goal is 
via a careful analysis of the Stokes profiles of some near-IR lines 
of \ion{Mn}{i} that we have measured with the Tenerife Infrared Polarimeter. 

The presence of all these small-scale magnetic fields in the quiet 
solar photosphere might have several important consequences for the 
overlying solar atmosphere, such as ubiquity of reconnecting current 
sheets and heating processes \citep[e.g.,][]{t2 Pr06}.
Therefore, it is now even more important to carry out detailed 
empirical investigations on the magnetism of the ``quiet'' solar 
chromosphere via a clever exploitation of a variety of subtle 
physical mechanisms by means of which a magnetic field can 
create and destroy spectral line polarization 
\citep[e.g.][]{t2 LL04}.

\acknowledgements
We are grateful to R. Casini (HAO) for carefully reviewing
our paper.
This research has been funded by the Spanish Ministerio de 
Educaci\'on y Ciencia through project AYA2004-05792.



\begin{thebibliography}

\bibitem[{{Anderson} \& {Athay}(1989)}]{t2 AA89}
{Anderson, I. S., \& Athay, R. G.} 1989, \apj, 346, 1010

\bibitem[{{Asensio Ramos}(2004)}]{t2 AR04}
{Asensio Ramos}, A. 2004, Ph.D.\ Thesis, University of La Laguna, La
Laguna, Tenerife, Spain
  
\bibitem[{{Asensio Ramos} \& {Trujillo Bueno}(2003)}]{t2 AT03}
{Asensio Ramos}, A., \& {Trujillo Bueno}, J. 2003, in ASP Conf. Ser.
Vol. 307, Solar Polarization 3, ed. J.~{Trujillo Bueno} \& 
J.~{S\'anchez Almeida} (San Francisco: ASP),~195
  
\bibitem[{Asensio Ramos \& Trujillo Bueno}(2005)]{t2 AT05}
Asensio Ramos, A., \& Trujillo Bueno, J. 2005, \apjl, 635, L112  

\bibitem[{Asplund et al.}(2000)]{t2 As00}
Asplund, M., Nordlund, \AA., Trampedach, R., Allende-Prieto, C.,
\& Stein, R. F. 2000, \aap, 359, 729

\bibitem[{Bellot Rubio} \& {Collados}(2003)]{t2 BC03}
Bellot Rubio, L., \& Collados, M. 2003, \aap, 406, 357

\bibitem[{{Berdyugina} \& {Fluri}(2004){Berdyugina}, \&
  {Fluri}}]{t2 BF04}
{Berdyugina}, S. V., \& {Fluri}, D. M. 2004, 
\aap, 417, 775

\bibitem[{{Berdyugina} {et~al.}(2002){Berdyugina}, {Stenflo}, \&
  {Gandorfer}}]{t2 Be02}
{Berdyugina}, S. V., {Stenflo}, J.~O., \& {Gandorfer}, A. 2002, 
\aap, 388, 1062

\bibitem[{Bommier \& Molodij}(2002)]{t2 BM02}
Bommier, V., \& Molodij, G. 2002, \aap, 381, 241

\bibitem[{Bommier et al.}(2005)]{t2 Bo05}
Bommier, V., Derouich, M., Landi Degl'Innocenti, E., Molodij, G., 
\& Sahal-Br\'echot, S. 2005, \aap, 432, 295

\bibitem[{Cattaneo}(1999)]{t2 Ca99}
Cattaneo, F. 1999, \apj, 515, L39

\bibitem[{de Wijn et al.}(2005)]{t2 Wi05}
De Wijn, A. G., Rutten, R. J., Haverkamp, E. M., \& 
S\"utterlin, P. 2005, \aap, 441, 1183

\bibitem[{Derouich}(2006)]{t2 De06}
Derouich, M. 2006, \aap, 449, 1

\bibitem[Derouich, Sahal-Br\'echot, \& Barklem(2005){Derouich et al.}]{t2 De05}
Derouich, M., Sahal-Br\'echot, S., \& Barklem, P. S. 2005, \aap, 434, 779

\bibitem[{Dom\'{\i}nguez Cerde\~na et~al.}(2003)Dom\'\i nguez Cerde\~na, Kneer, 
\& S\'anchez Almeida]{t2 DC03}
Dom\'\i nguez Cerde\~na, I., Kneer, F. \& S\'anchez Almeida, J. 
2003, \apj, 582, L55

\bibitem[{Dom\'{\i}nguez Cerde\~na et~al.}(2006)Dom\'\i nguez Cerde\~na, 
S\'anchez Almeida, \& Kneer]{t2 DC06}
Dom\'\i nguez Cerde\~na, I., S\'anchez Almeida, J., 
\& Kneer, F. 2006, \apj, 636, 496

\bibitem[{Faurobert} \& {Arnaud}(2002)]{t2 FA02}
Faurobert, \& M., Arnaud, J. 
2002, \aap, 382, L17

\bibitem[{Faurobert} \& Arnaud(2003)]{t2 FA03}
Faurobert, \& M., Arnaud, J. 
2003, \aap, 412, 555

\bibitem[{Faurobert} et al.(2001)]{t2 Fa01}
Faurobert, M., Arnaud, J., Vigneau, J., 
\& Frisch, H. 2001, \aap, 378, 627

\bibitem[{Faurobert-Scholl et al.}(1995)]{t2 FS95}
Faurobert-Scholl, M., Feautrier, N., Machefert, F., Petrovay, K., 
\& Spielfiedel, A. 1995, \aap, 298, 289

\bibitem[{Gandorfer}(2000)]{t2 Ga00}
Gandorfer, A. 2000, The Second Solar Spectrum: A High Spectral 
Resolution Polarimetric Survey of Scattering Polarization at the 
Solar Limb in Graphical Representation. Vol. I: 4625\,\AA\ to 6995\,\AA\ 
(Zurich: vdf ETH) 

\bibitem[{Gandorfer}(2002)]{t2 Ga02}
Gandorfer, A. 2002, The Second Solar Spectrum: A High Spectral 
Resolution Polarimetric Survey of Scattering Polarization at the 
Solar Limb in Graphical Representation. Vol. II: 3910\,\AA\ to 4630\,\AA\
(Zurich: vdf ETH) 

\bibitem[{Gandorfer}(2003)]{t2 Ga03}
Gandorfer, A. 2003, 
in ASP Conf. Ser. Vol. 307, Solar Polarization 3, ed. 
J. Trujillo Bueno \& J. S\'anchez Almeida (San Francisco: ASP),~399

\bibitem[{Herzberg}(1950)]{t2 He50}
Herzberg, G. H. 1950, Molecular Spectra and Molecular Structure: 
I. Spectra of Diatomic Molecules, 2nd edn. (Princeton: Van Nostrand)

\bibitem[{Keller et al.}(1994)]{t2 Ke94}
Keller, C. U., Deubner, F. L., Egger, U., Fleck, B., \& Povel, H. P. 1994,
\aap, 286, 626

\bibitem[{Khomenko}(2006)]{t2 Kh06}
Khomenko, E. V. 2006, in ASP Conf. Ser. Vol. in press,
Solar MHD: Theory and Observations: a High Spatial Resolution 
Perspective, ed. J. Leibacher, H. Uitenbroek \& R. F. Stein
(San Francisco: ASP)

\bibitem[{Khomenko \& Collados}(2006)]{t2 KC06}
Khomenko, E. V., \& Collados, M. 2006, these proceedings

\bibitem[{Khomenko et al.}(2003)]{t2 Kh03}
Khomenko, E. V., Collados, M., Solanki, S. K., Lagg, A., 
\& Trujillo Bueno, J. 2003, \aap, 408, 1115

\bibitem[{Khomenko et al.}(2005)]{t2 Kh05}
Khomenko, E. V., Shelyag, S., Solanki, S. K., \& V\"ogler, A. 2005, \aap,
442, 1059

\bibitem[{Landi Degl'Innocenti}(2003)]{t2 La03}
Landi Degl'Innocenti, E. 2003, in ASP Conf. Ser. Vol. 307,
Solar Polarization 3, ed. J. Trujillo Bueno \& J. S\'anchez Almeida
(San Francisco: ASP),~164

\bibitem[{Landi Degl'Innocenti}(2006)]{t2 La06}
Landi Degl'Innocenti, E. 2006, \aap, in press

\bibitem[{Landi Degl'Innocenti}(2006)]{t2 La06b}
Landi Degl'Innocenti, E. 2006, these proceedings

\bibitem[{Landi Degl'Innocenti \& Landolfi}(2004)]{t2 LL04}
Landi Degl'Innocenti, E., \& Landolfi, M. 2004, Polarization in 
Spectral Lines (Dordrecht: Kluwer)

\bibitem[{Lin \& Rimmele}(1999)]{t2 LR99}
Lin, H., \& Rimmele, T. 1999, \apj, 514, 448

\bibitem[{Lites}(2002)]{t2 Li02}
Lites, B. W. 2002, \apj, 573, 431

\bibitem[{Lites \& Socas-Navarro}(2004)]{t2 LS04}
Lites, B. W., \& Socas-Navarro, H. 2004, \apj, 613, 600

\bibitem[{L\'opez Ariste et al.}(2002)L\'opez Ariste, Tomczyk, 
\& Casini]{t2 LA02}
L\'opez Ariste, A., Tomczyk, S., \& Casini, R. 2002, \apj, 580, 519

\bibitem[{Manso Sainz et al.}(2004)Manso Sainz, Landi Degl'Innocenti, 
\& Trujillo Bueno]{t2 MS04}
Manso Sainz, R., Landi Degl'Innocenti, E., \& Trujillo Bueno, J. 2004, 
\apj, 614, L89

\bibitem[{Mart\'{\i}nez Gonzalez et al.}(2006)Mart\'{\i}nez 
Gonz\'alez, Collados, \& Ruiz Cobo]{t2 MG06}
Mart\'{\i}nez Gonz\'alez, M., Collados, M., \& Ruiz Cobo, B. 2006, these
proceedings

\bibitem[{Pierce}(1968)]{t2 Pi68}
Pierce, A. K. 1968 \apjs, 17, 1

\bibitem[{Priest}(2006)]{t2 Pr06}
Priest, E. 2006, in The Many Scales in the Universe, ed. J. C. 
del Toro Iniesta (Dordrecht: Kluwer), 197.

\bibitem[{S\'anchez Almeida}(2005)]{t2 SA05}
S\'anchez Almeida, J., 2005, \aap, 438, 727

\bibitem[{S\'anchez Almeida \& Lites}(2000)]{t2 SL00}
S\'anchez Almeida, J., \& Lites, B. W. 2000, \apj, 532, 1215

\bibitem[{S\'anchez Almeida et al.}(1996)]{t2 SA96}
S\'anchez Almeida, J., Landi Degl'Innocenti, E., Mart\'{\i}nez Pillet, V., 
\& Lites, B. W. 1996, \apj, 466, 537

\bibitem[S\'anchez Almeida, Emonet, 
\& Cattaneo(2003a){S\'anchez Almeida et al.}]{t2 SA03a}
S\'anchez Almeida, J., Emonet, T., \& Cattaneo, F. 2003a, \apj, 585, 536

\bibitem[{S\'anchez Almeida et al.}(2003b)S\'anchez Almeida, 
Dom\'{\i}nguez Cerde\~na, \& Kneer]{t2 SA03b}
S\'anchez Almeida, J., Dom\'{\i}nguez Cerde\~na, I., \& Kneer, F. 2003b, \apj, 597, L177

\bibitem[{S\'anchez Almeida et al.}(2004)]{t2 SA04}
S\'anchez Almeida, J., M\'arquez, I., Bonet, J. A.,
Dom\'{\i}nguez Cerde\~na, I., \& Muller, R. 2004, \apj, 609, L91

\bibitem[{Schrijver} \& Title(2003)]{t2 ScT03}
Schrijver, C. J., \& Title, A. 2003, \apj, 597, L165

\bibitem[{Shchukina \& Trujillo Bueno}(2001)]{t2 ST01}
Shchukina, N., \& Trujillo Bueno, J. 2001, \apj, 550, 970

\bibitem[{Shchukina \& Trujillo Bueno}(2003)]{t2 ST03}
Shchukina, N., \& Trujillo Bueno, J. 2003, in ASP Conf. Ser. Vol. 307,
Solar Polarization 3, ed. J. Trujillo Bueno \& J. S\'anchez Almeida
(San Francisco: ASP),~336

\bibitem[{Shchukina \& Trujillo Bueno}(2006)]{t2 ST06}
Shchukina, N., \& Trujillo Bueno, J. 2006, in preparation

\bibitem[{Socas-Navarro \& Lites}(2004)]{t2 SL04}
Socas-Navarro, H., \& Lites, B. W. 2004, \apj, 616, 587

\bibitem[{Socas-Navarro \& S\'anchez Almeida}(2002)]{t2 SS02}
Socas-Navarro, H., \& S\'anchez Almeida, J. 2002, \apj, 565, 1323

\bibitem[{Socas-Navarro \& S\'anchez Almeida}(2003)]{t2 SS03}
Socas-Navarro, H., \& S\'anchez Almeida, J. 2003, \apj, 593, 581

\bibitem[Socas-Navarro, Mart\'\i nez Pillet, \& Lites(2004){Socas-Navarro et al.}]{t2 SN04}
Socas-Navarro, H., Mart\'\i nez Pillet, V., \& Lites, B. W. 2004, 
\apj, 611, 1139

\bibitem[{Stein \& Nordlund}(2003)]{t2 SN03}
Stein, R. F., \& Nordlund, \AA. 2003, in IAU Symp. 210, Modeling of 
Stellar Atmospheres, ed. N. E. Piskunov, W. W. Weiss \& D. F. Gray 
(San Francisco: ASP), 169

\bibitem[{Stenflo}(1973)]{t2 St73}
Stenflo, J. O. 1973, \solphys, 32, 41

\bibitem[{Stenflo}(1982)]{t2 St82}
Stenflo, J. O. 1982, \solphys, 80, 209

\bibitem[{Stenflo}(1994)]{t2 St94}
Stenflo, J. O. 1994, Solar Magnetic Fields: Polarized Radiation
Diagnostics (Dordrecht: Kluwer)

\bibitem[{Stenflo}(2003)]{t2 Sten03}
Stenflo, J. O. 2003, in ASP Conf. Ser. Vol. 307,
Solar Polarization 3, ed. J. Trujillo Bueno \& J. S\'anchez Almeida
(San Francisco: ASP),~385

%
%

\bibitem[{Stenflo \& Holzreuter}(2003)]{t2 SH03}
Stenflo, J. O., \& Holzreuter, R. 2003, in ASP Conf. Ser. Vol. 286,
Current Theoretical Models and Future High Resolution Solar 
Observations: Preparing for ATST, ed. A. A. Pevtsov \& H. Uitenbroek
(San Francisco: ASP),~169

\bibitem[{Stenflo \& Keller}(1997)]{t2 SK97}
Stenflo, J. O., \& Keller, C. U. 1997, \aap, 321, 927

\bibitem[{Stenflo et al.}(1997)]{t2 St97}
Stenflo, J. O., Bianda, M., Keller, C. U., \& Solanki, S. K. 1997,
\aap, 322, 985

\bibitem[{Thelen \& Cattaneo}(2000)]{t2 TC00}
Thelen, J. C., \& Cattaneo, F. 2000, \mnras, 315, L13

\bibitem[{Trujillo Bueno}(2001)]{t2 TB01a}
Trujillo Bueno, J. 2001, in ASP Conf. Ser. Vol. 236, Advanced Solar 
Polarimetry: Theory, Observation and Instrumentation, ed. M. Sigwarth
(San Francisco: ASP),~161

\bibitem[{Trujillo Bueno}(2003a)]{t2 TB03a}
Trujillo Bueno, J. 2003a, in ASP Conf. Ser. Vol. 307, Solar 
Polarization 3, ed. J. Trujillo Bueno \& J. S\'anchez Almeida (San
Francisco: ASP),~407

\bibitem[{Trujillo Bueno}(2003b)]{t2 TB03b}
Trujillo Bueno, J. 2003b, in ASP Conf. Ser. Vol. 288, Stellar Atmosphere 
Modeling, ed. I. Hubeny, D. Mihalas \& K. Werner (San Francisco: ASP),~551

\bibitem[{Trujillo Bueno}(2003c)]{t2 TB03c}
Trujillo Bueno, J. 2003c, in IAU Symp. 210, Modelling of Stellar 
Atmospheres, ed. N. Piskunov, W. W. Weiss \& D. F. Gray 
(San Francisco: ASP),~243

\bibitem[{Trujillo Bueno \& Manso Sainz}(1999)]{t2 TM99}
Trujillo Bueno, J., \& Manso Sainz, R. 1999, \apj, 516, 436

\bibitem[{Trujillo Bueno et al.}(2001)]{t2 TB01b}
Trujillo Bueno, J., Collados, M., Paletou, F., \& Molodij, G. 2001,
in ASP Conf. Ser. Vol. 236, Advanced Solar Polarimetry: Theory, 
Observation and Instrumentation, ed. M. Sigwarth (San Francisco: ASP),~141

\bibitem[{Trujillo Bueno et al.}(2004)Trujillo Bueno, Shchukina, 
\& Asensio Ramos]{t2 TB04}
Trujillo Bueno, J., Shchukina, N., \& Asensio Ramos, A. 2004, \nat, 
430, 326

\bibitem[{V\"ogler}(2003)]{t2 Vo03}
V\"ogler, A. 2003, Ph.D.\ Thesis, University of G\"ottingen, G\"ottingen,
Germany

\bibitem[{Weck} et al.(2003)]{t2 We03}
Weck, P. F., Schweitzer, A., Stancil, P. C., Hauschildt, P. H., 
\& Kirby, K. 2003, \apj, 582, 1059

\bibitem[{Woo}(2006)]{t2 Wo06}
Woo, R. 2006, \apj, 639, L95

\end{thebibliography}
\end{document}